\input{epsf}

\documentclass[12pt, draftclsnofoot, peerreview, onecolumn]{IEEEtran}
\usepackage{epsf}
\usepackage{graphicx}
\usepackage[cmex10]{amsmath}
\usepackage{amsthm}
\usepackage{amssymb}
\usepackage{epsfig,latexsym,amsmath,epsf,amssymb,amsfonts}
\usepackage{algorithm, algorithmic}
\usepackage{placeins}
\usepackage{cite}
\usepackage{comment} 
\usepackage{subfigure}
\usepackage{lipsum}

\usepackage{bbm}

\renewcommand{\theequation}{\arabic{equation}}
\newtheorem{Definition}{\hskip 0pt Definition}
\newtheorem{Theorem}{\hskip 0pt Theorem}
\newtheorem{Lemma}{\hskip 0pt Lemma}
\newtheorem{Proposition}{\hskip 0pt Proposition}

\newtheorem{Remark}{\hskip 0pt Remark}

\begin{document}
%
\title{Learning for Robust Routing Based on Stochastic Game in Cognitive Radio Networks}

\author{
\IEEEauthorblockN{Wenbo Wang ~\IEEEmembership{Student Member,~IEEE,}
Andres Kwasinski ~\IEEEmembership{Senior Member,~IEEE,} 
Dusit Niyato ~\IEEEmembership{Senior Member,~IEEE,}  and
Zhu Han ~\IEEEmembership{Fellow,~IEEE}\vspace*{-4mm}}

\thanks{Wenbo Wang and Andres Kwasinski are with the Department of Computer Engineering, Rochester Institute of Technology, Rochester, NY 
14623 USA (email: wxw4213@rit.edu, axkeec@rit.edu).}
\thanks{Dusit Niyato is with the School of Computer Engineering, Nanyang Technological University, Singapore 639798
(email: dniyato@ntu.edu.sg).}
\thanks{Zhu Han is with the Department of Electrical and Computer Engineering as well as the Department of Computer Science, University of Houston, TX 77004 USA
(email: zhan2@uh.edu).}\vspace*{-4mm}}


\maketitle

\begin{abstract}
This paper studies the problem of robust spectrum-aware routing in a multi-hop, multi-channel Cognitive Radio Network (CRN) with the presence of malicious nodes in the secondary 
network. The proposed routing scheme models the interaction among the Secondary Users (SUs) as a stochastic game. By allowing the backward propagation of the path utility information
from the next-hop nodes, the stochastic routing game is decomposed into a series of stage games. The best-response policies are learned through the process of smooth fictitious 
play, which is guaranteed to converge without flooding of the information about the local utilities and behaviors. To address the problem of mixed insider attacks with both 
routing-toward-primary and sink-hole attacks, the trustworthiness of the neighbor nodes is evaluated through a multi-arm bandit process for each SU. The
simulation results show that the proposed routing algorithm is able to enforce the cooperation of the malicious SUs and reduce the negative impact of the attacks on the routing 
selection process.
\end{abstract}

\vspace*{-2mm}

\section{Introduction}
In Cognitive Radio Networks (CRNs), Dynamic Spectrum Access (DSA) policies require Secondary Users (SUs) to opportunistically access idle channels which are temporarily unused 
by the Primary Users (PUs). Although being considered an efficient way of spectrum utilization \cite{4796930}, with DSA, SUs do not have channels which are always available to 
access. As a result, 
new coupling between the PHY-MAC layers and upper layer protocols arises. At the network layer, a routing protocol is thus expected to explicitly address the impact of unstable 
channels on the topology and the link performance of the secondary network. Generally, routing problems in CRNs exhibits a certain level of similarity to routing problems 
in multi-channel ad hoc networks \cite{Cesana2011228}. However, routing in CRNs faces a number of new, different challenges \cite{Cesana2011228, 6599059}:
\begin{itemize}
 \item[(i)] spectrum awareness: timely adaptation to the dynamic change of the channel availability due to DSA, and 
 \item[(ii)] self-organization: proper route configuration with the limited/heterogeneous level of channel resource knowledge.
\end{itemize}

Due to the information uncertainty or locality caused by DSA, a distributed route discovery process in CRNs also tends to be more vulnerable to the insider attacks than that in 
conventional ad-hoc networks. As indicated by \cite{5742780, liu2010cognitive}, an attacker in CRNs exploits the following characteristics of DSA schemes:
(i) vulnerabilities due to information locality with respect to sensing and reporting of spectrum states, and (ii) the imperfect knowledge of SUs 
about the time-varying PU channels. Since the accuracy of the channel state information directly affects the performance of the DSA schemes in CRNs, most of the 
identified attacks in CRNs target at channel state information distortion for attacking the PHY-MAC layer protocols \cite{4413138, 6138837, 5738229}. Similarly, 
it is now possible for routing attackers to bypass the network-layer vulnerabilities used by traditional routing attack schemes and only need to distort the channel 
detection/access information in the DSA mechanism to disrupt the routing process.

In this paper, we study the routing mechanism in a multi-hop, multi-channel CRN and address the challenges of spectrum awareness, information locality and routing security as a 
joint problem. We consider limited spectrum sensing abilities of each SU in a real-world situation. We also consider the presence of malicious SUs which 
can apply sophisticated attacks by combining different methods of attacks including Sink-Hole (SH) and Routing-toward-Primary-User (RPU). In order to tackle 
the routing-under-attack problem for multiple flows, we formulate the joint channel-relay selection process of the SUs as a stochastic game. We propose a distributed, adaptive 
channel-relay selection scheme for SUs to learn their routing strategies with only limited amount of information exchange. To defend the SH attack with 
information distortion, we model the routing performance evaluation process as a Multi-Arm Bandit (MAB) problem and use the estimated arm-selection probability as an indicator 
of the neighbor trustworthiness. The proposed routing scheme is featured as a self-organized strategy-learning process in a series of single-state repeated games.
Neither the a-priori channel activity model nor information flooding among the SUs is required for implementing the learning scheme. 
\vspace{-10.5pt}
\section{Related Work}
\subsubsection{Routing in CRNs}
The solutions to the routing problems in CRNs are usually featured by a cross-layer design that directly integrates channel sensing and MAC operations into the routing 
protocols. These routing protocols may vary significantly due to different assumptions on the PU activity model and DSA mechanisms. Such variation is usually reflected by 
differences in the selection of link metrics and routing scheme types (e.g., reactive and proactive). With respect to the different channel occupation models (e.g., 
underlay vs. overlay/interweaving), the link metrics may be designed in different ways. For overlay/interweaving CRNs, many studies designed the routing mechanism based on 
a snapshot of the channel dynamics \cite{4518957,5935140,4413147, 6178255, 6419858}. In these studies, delay-based link quality metrics were proposed based on the collision map 
for the SUs over the PU channels. For underlay CRNs, the link quality metric may be designed based on the link capacity as a function of the interference to the PUs 
\cite{5430958}. For both groups of solutions, routing schemes were usually designed in a time-slotted manner to analyze and optimize the impact of DSA mechanisms on the route performance.
If complete information on the channel states and local routing decisions is assumed, the routing problem is usually formulated as an optimization programming 
problem (e.g., convex or integer programming) and solved with a centralized route scheduler \cite{5935140, 4413147, 6178255, 6419858, 5430958}. 

In contrast to routing mechanisms using instantaneous collision maps, a number of works designed their link quality metric based on an a-priori probabilistic channel dynamic 
model \cite{6225388, 6331686, Yang:2008:LCB:1413939.1413945}. Since the impact of DSA schemes on the link performance is reflected by the 
stochastic channel activity model, it is possible for the secondary network to treat the routing problem in CRNs as a routing problem in conventional ad-hoc networks.
As a result, we can adopt 
existing protocols (e.g., link state routing \cite{6225388}, AODV \cite{Yang:2008:LCB:1413939.1413945} and RPL \cite{6857431}) with little modification. The advantage 
of such an approach is that it provides a way of reflecting the channel dynamics in the probabilistic link metrics based on the stochastic channel activity model. Hence, 
the routing protocols does not need to consider the instantaneous impact of the PHY-MAC layers. Since no collision map or route scheduler is needed, such an
approach is more appropriate for designing a distributed routing mechanism. However, many of these distributed routing mechanisms only provide a
heuristic routing solution. Also, it is often unrealistic to assume an a-priori channel activity model in practical scenarios and the applicability of deploying the 
aforementioned routing mechanisms may be limited.

In practice, the channel dynamics may exhibit heterogeneous characteristics with respect to the geolocation. In addition, SUs may have limited capability of acquiring 
information about the channel states and their neighbors' behaviors. Consequently, game theoretic analysis have become the focus of CRN routing protocol design, since it 
can efficiently solve the distributed control problems with constraints on the information exchange. Game-based routing solutions can be found 
in the studies on spectrum-aware, multi-flow routing \cite{4600227, 6331690} and traffic engineering \cite{5072224} in CRNs. In these studies, the SUs are assumed to be 
non-malicious and honest in sharing information, and the model of repeated (noncooperative) games is usually applied. The cooperation among the SUs is implicitly enforced through
repeatedly playing the game and the performance of a route is ensured by the value of the game.

\subsubsection{Security Issues for Routing in CRNs}
In the literature, most of the studies on security problems in routing protocols target conventional ad-hoc networks \cite{4396947, Xiao2007attack}. In these studies, 
the main focus is to prevent information distortion (e.g., with public-key distribution \cite{1306970}) or to identify the attackers with limited traffic monitoring (e.g., 
\cite{Marti:2000:MRM:345910:345955, 6997490}). When game theoretic solutions are adopted, the interaction between the honest and 
malicious nodes is typically modeled as a constant/zero-sum game and solved by obtaining the minimax equilibrium strategies in the game (e.g., \cite{1209210, 4288123}). 

There are relatively few works on the secured routing protocol design in CRNs. Among them, most of the studies are confined to handling the jamming attacks
or PUE attacks which distort the quality of an established link between the legitimated (normal) SUs (e.g., \cite{6133879}). A more sophisticated routing attacks in CRNs recently 
identified is the Routing-toward-Primary-User (RPU) attack in multi-hop, overlay CRNs \cite{6226410}. Unlike the PHY-MAC-layer dominated attacks, the
RPU attack exploits the geographical heterogeneity of PU activities and tunnels the traffic to the SUs in the footprint of the PU transmission. An RPU attacker emulates a 
combined attacking mechanism of both the Sink-Hole (SH) attack \cite{4396947} and the Selective-Forwarding (SF) attack \cite{4396947}. However, the RPU-caused packet drop/delay 
is not directly due to the network layer operation, but due to the collisions with PU transmissions on the PHY-MAC layers.

The paper is organized as follows. Section \ref{sec_modeling} describes the models of the PU activities and the SU behaviors. Based on these models, a spectrum-aware 
link quality metric is proposed to reflect the impact of the channel state dynamics on the routing process. In Section \ref{sec_routing_game}, the multi-flow routing 
process in the secondary network is formulated as a layered average-reward stochastic game and then is shown to be equivalent to a group of single-state repeated games. In 
Section \ref{sec_learning_truthful_game}
and Section \ref{sec_learning_sophisticated_game}, am adaptive strategy-learning mechanism and a trustworthiness-evaluation mechanism are proposed for the normal SUs to seek the 
best-response routing
strategies against the attackers with limited information exchange. The simulation results are provided in Section \ref{sec_simulation} to demonstrate the Effectiveness of the 
proposed routing mechanism. Section \ref{sec_conclusion} concludes the contribution of this paper.
\vspace{-10.5pt}
\section{Network Model}
\label{sec_modeling}
We consider a multi-hop CRN that interweaves upon $K$ orthogonal PU channels. The normal SUs abide by the interweaving DSA rule and establish links over the
temporarily free PU channels. The nodes in the CRN are divided into three types: the source SUs, sink SUs and relay SUs. We consider that the relay SUs do not generate 
packets and only forward the received packets to their neighbors. Among the relay SUs, some malicious nodes
adopt RPU-like attacks to cause delay to the traffic as much as possible.
\vspace{-10.5pt}
\subsection{Dynamic Spectrum Access Model}
\label{sub_sec_dsa}
Based on the empirical study of the PU channel occupation time in \cite{4413143}, we assume that the PU activities over each channel can be modeled as an independent 
continuous-time Markov process with the binary states \emph{Idle} (`0') and \emph{Busy} (`1').  For a channel $k$, we assume that $\lambda_k^{-1}$ and $\mu_k^{-1}$ are the 
mean holding times for states \emph{Idle} and \emph{Busy}, respectively. Then, the corresponding transition matrix is given by \cite{4413143} as follows: 
\begin{equation}
  \label{eq_transition_matrix}
  \mathbf{P}_k(t)\!=\!\frac{1}{\lambda_k\!+\!\mu_k}\!
  \begin{pmatrix}
    \mu_k\!+\!{\lambda_k}e^{\!-\!(\lambda_k\!+\!\mu_k)t}\!&\!\lambda_k-\lambda_k e^{\!-\!(\lambda_k\!+\!\mu_k)t}\\
    \mu_k\!-\!\mu_k e^{\!-\!(\lambda_k\!+\!\mu_k)t}\!&\!\lambda_k\!+\!\mu_k e^{\!-\!(\lambda_k\!+\!\mu_k)t}
  \end{pmatrix}.
\end{equation}
Since in practical scenarios the PU activities are usually geographically different, we assume that the CRN can be geographically divided into a set of non-overlapping,
independent spectrum activity clusters according to the local PU activities. For conciseness, we consider a snapshot of the network, during which the cluster 
topology remains unchanged. The cluster topology can be managed in a similar way to the CogMesh protocol \cite{4221491} by trustworthy cluster heads. The 
cluster heads maintain the cluster formation through message exchange with neighbor nodes using a dedicated control channel.

\begin{figure}
 \begin{minipage}{.49\textwidth}
  \centering
  \includegraphics[width=1.0\textwidth]{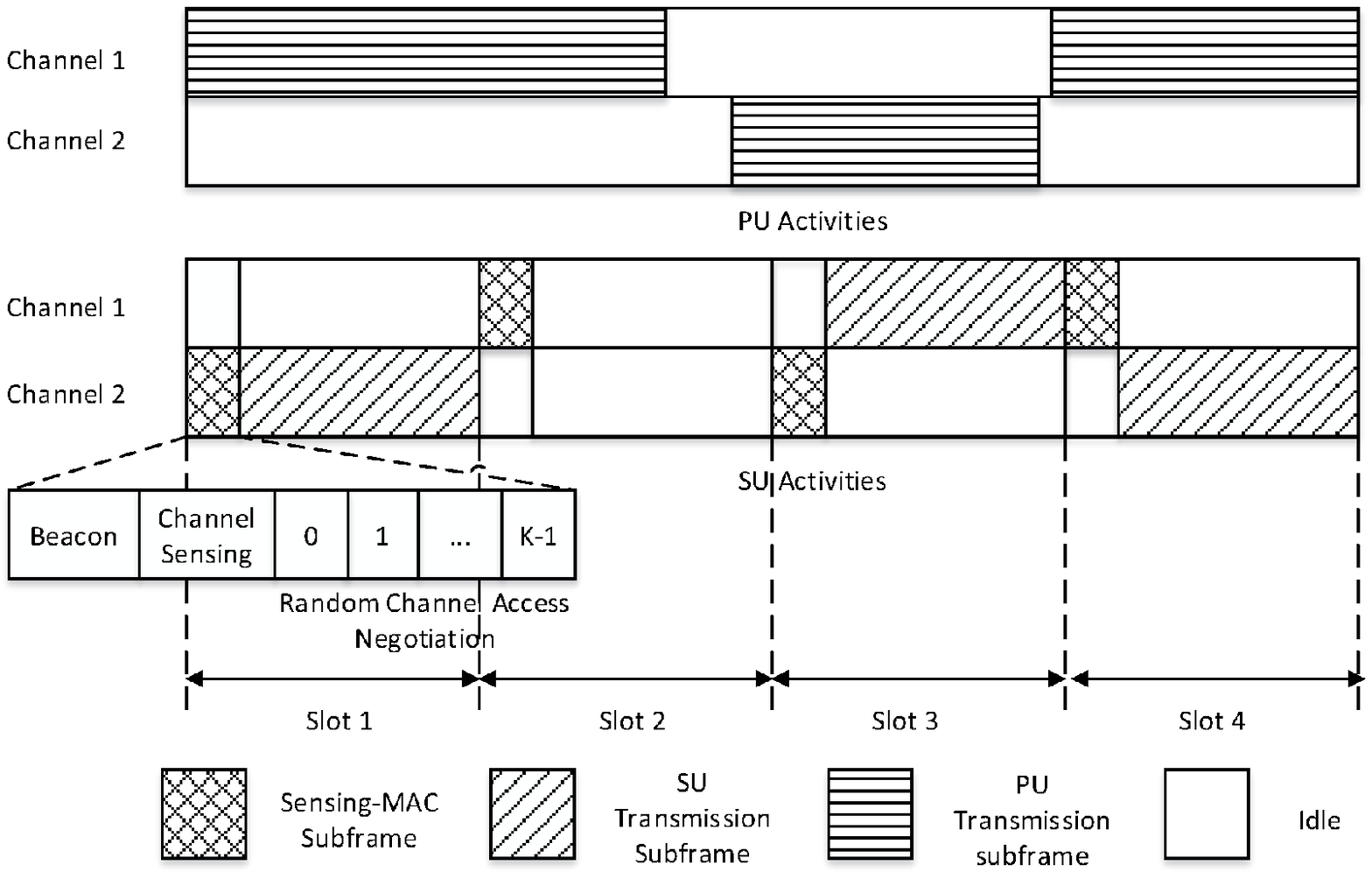}
  \caption{Radio activities in a two-channel CRN with coordinated periodic sensing in the secondary network.}
  \label{fig1}
 \end{minipage}
 \begin{minipage}{.49\textwidth}
  \centering
  \includegraphics[width=1.0\textwidth]{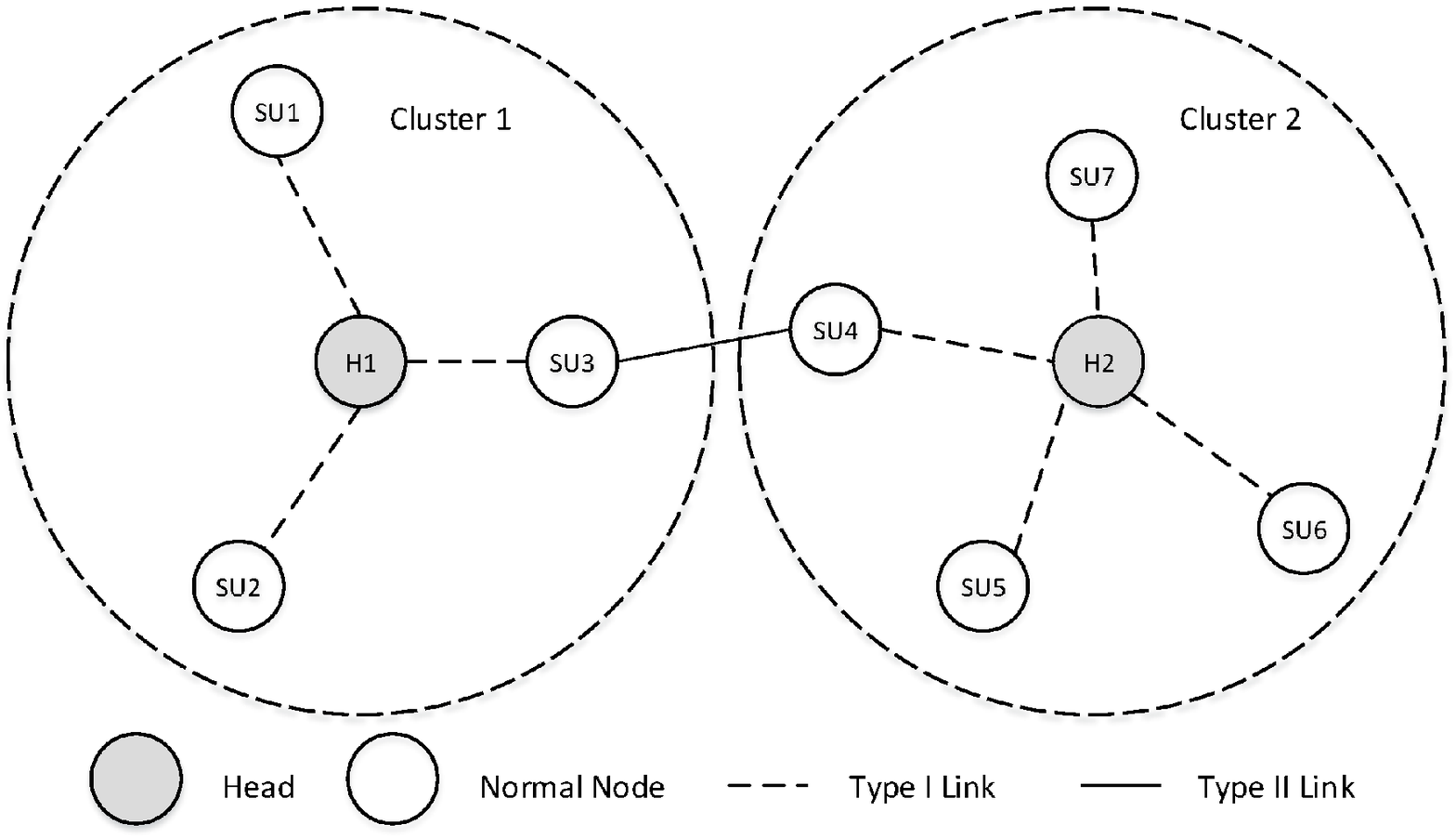}
  \caption{The SU links in a CRN of two clusters.}
  \label{fig2}
 \end{minipage}
\end{figure}

We assume that SUs access the PU channels in a time slotted manner. Due to the practical limit on the number of radio interfaces in each SU, we assume that an SU can only sense 
one PU channel during one sensing slot. To reduce the detection error, SUs in the same cluster sense the PU channels following a round-robin schedule in an ascending order of
the channel indices (Figure \ref{fig1}). The sensing results from the SUs in the same cluster are aggregated by the cluster head \cite{4796930}. 
We assume that the detection error is negligible with aggregated sensing. For a cluster, the state of channel $i$, $i\!\in\!\mathcal{K}\!=\!\{0,\ldots,K\!-\!1\}$, 
is updated only when 
$i\!=\!(n\!\mod\!K)$ at slot $n$. For channels $k$, $k\!\ne\!i$, the SUs in the cluster keep the most recent sensing result at slot $\phi_{k}(n)\!=\!n\!-\![(K\!+\!i\!-\!k)\!\mod
\!K]$ as their estimated state. For cluster $q$, let $\mathbf{o}^q(n)\!=\![o^q_0(n), o^q_1(n), \ldots, o^q_{K-1}(n)]^{\mathsf{T}}$ denote the vector of estimated channel states at slot $n$, 
and $\mathbf{s}^q(n)\!=\![s^q_0(n), s^q_1(n), \ldots, s^q_{K-1}(n)]^{\mathsf{T}}$ denote the real channel state vector. According to \cite{4410464}, the process 
$\mathbf{o}^j(n)$ is an irreducible, periodic, discrete-time Markov chain. Let $i'\!=\!(n\!+\!1)\!\mod\!K$ be the channel sensed at slot $(n+1)$, then the transition probability of 
$\mathbf{o}^q(n)$ can be obtained based on (\ref{eq_transition_matrix}) as:
\begin{equation}
  \label{eq_sensing_state}
  \begin{array}{ll}
  P(\mathbf{o}_j^q(n\!+\!1)\!=\!{s}'|\mathbf{o}_j^q(n)\!=\!{s})\!=\!\left\{\!
  \begin{array}{ll}
   \left[\mathbf{P}^q_j(KT)\right]_{({s},{s}')},\textrm{ if } j=i',\\
   0, \qquad\qquad\quad\textrm{ otherwise},
  \end{array}\right.
  \end{array}
\end{equation}
where $\mathbf{P}^q_j$ is the transition matrix of channel $j$ in cluster $q$, $T$ is the slot length and $[\mathbf{P}^q_j(KT)]_{({s},{s}')}$ is the element of $\mathbf{P}^q_j$ transiting
from $s$ to $s'$.
\vspace{-10.5pt}
\subsection{Impact of Node Behavior on Link Quality}
\label{sub_sec_node_behavior}
Let $\mathcal{N}_i$ denote the set of one-hop neighbors of an SU $i$ (including $i$). We assume that in slot $n$, SU $i$ can freely choose its target relay SU in the neighborhood
and target channel among the PU channels for data forwarding, if no constraint on SU behaviors is presented. We denote such an action by the action vector $a_i(n)\!=\!(j,k)$, 
where $j\!\in\!\mathcal{N}_i\backslash\{i\}$ and $k\!\in\!\mathcal{K}$. In a multi-hop CRN, it is natural to consider that the more hops used for packet forwarding, the larger 
total delay the path has. To enforce that packets are forwarded toward the sink SUs and no cyclic path is formed, each SU is able to exchange its geographical information with 
its neighbors. Using the geographical information of the neighbor SUs, we introduce the distance advancement metric of a
relay toward its sink \cite{6331686} to help evaluate the link quality. Let $L$ denote the sink SU, the distance advancement of SU $i$ by choosing $a_{i}=(j,k)$ is 
defined as the reduction of distance from SU $i$ to SU ${L}$ when routing via SU $j$:
\begin{equation}
 \label{eq_distance_advancement}
  A_{i}(a_i)=D(i,L)-D(j,L),
\end{equation}
where $D(i,j)$ is the Euclidean distances between SUs $i$ and $j$. Based on the relay distance advancement in (\ref{eq_distance_advancement}), we impose the rule 
that an SU is forbidden to select relays that produce negative distance advancement. Then for SU $i$, the set of candidate actions for channel-relay selection $a_{i}=(j,k)$
is defined by $\{j:j\in\mathcal{N}_i\backslash\{i\}, A_i(j)\ge 0\}$.

Let $(i,j)_k$ denote the link formed from SU $i$ to SU $j$ over channel $k$ when SU $i$ takes action $a_i=(j,k)$.
According to the rule of DSA interweaving, link $(i,j)_k$ is accessible only when channel $k$ is free for both SUs.
Let $q(i)$ denote the spectrum activity cluster that SU $i$ is in, then link $(i,j)_k$ can be classified into two types as shown in Figure \ref{fig2}:
\begin{itemize}
 \item Type I: $i$ and $j$ are in the same cluster: $q(i)=q(j)$.
 \item Type II: $i$ and $j$ are in different clusters: $q(i)\ne q(j)$.
\end{itemize}
For Type I links, we only need to consider the channel state of one cluster, while for Type II links it is necessary to consider the joint channel state evolution of the two 
involved clusters.

We consider that the quality of link $(i,j)_k$ is measured based on the Effective Transmission Time (ETT) \cite{Draves:2004:RMM:1023720.1023732}.  When the link is stable, the 
ETT over link $(i,j)_k$ can be measured as:
\begin{equation}
 \label{eq_ett}
 d^{\textrm{ETT}}_{(i,j)_k}=\frac{L}{R(1-P_{e}(k))},
\end{equation}
where $L$ is the packet length, $R$ is the transmit rate and $P_e(e)$ is the packet error rate due to the physical layer error over channel $k$. When lacking stable 
channels, it is necessary to explicitly reflect in the link quality metric the impact of the DSA mechanism and MAC protocol. We note from (\ref{eq_sensing_state}) that due 
to the imperfect knowledge on channel states, a transmission failure may occur in the secondary network even when the current channel state vector 
indicates that channel $k$ is free. Therefore, in order to determine the accessibility of link $(i,j)_k$, it is necessary to consider the conditional probability for channel $k$ 
to be \emph{Idle}  during slot $n$ given the observed state vectors of clusters $q(i)$ and $q(j)$ at the beginning of slot $n$. Based on (\ref{eq_transition_matrix}), the  
probability for channel $k$ to be \emph{Idle}  for a period $\tau$ from the beginning of slot $n$ in cluster $q(i)$ can be calculated as:
\begin{eqnarray}
 \label{eq_conditional_pr_free_channel}
 \begin{array}{ll}
  P_k^{q(i)}(\tau,\mathbf{o}^{q(i)}_k(n))
  =P_k^{q(i)}\left(\mathbf{s}^{q(i)}_k(nT\!+\!\tau)\!=\!0|\mathbf{o}^{q(i)}_k(n)=\mathbf{s}^{q(i)}_k\left(\phi_k(n)T\right)\right)\\
  =\!e^{-\lambda^{q(i)}_k\tau}\left[\mathbf{P}^{q(i)}_k((n-\phi_k(n))T)\right]_{\left(\mathbf{s}^{q(i)}_k(\phi_k(n)T),0\right)},
 \end{array}
\end{eqnarray}
where $e^{-\lambda^{q(i)}_k\tau}$ is the probability for the channel to remain idle for time $\tau$ since the beginning of slot $n$, and $\left[\mathbf{P}^{q(i)}_k((n-\phi_k(n))T)
\right]_{\left(\mathbf{s}^{q(i)}_k(\phi_k(n)T),0\right)}$ is obtained from (\ref{eq_transition_matrix}).

The link availability probability for $(i,j)_k$ at slot $n$ depends on the probability of channel $k$ staying idle at both end SUs. Based on our discussion of the link
type, the probability of channel $k$ being available for link $(i,j)_k$ during slot $n$ can be expressed as:
\begin{eqnarray}
 \label{eq_prob_free_link}
  \begin{array}{ll}
 P_k^{i,j}(\mathbf{o}(n))\!=\!P_k^{i,j}(\mathbf{o}^{q(i)}_k(n),\mathbf{o}^{q(j)}_k(n))\!=\!
 \left\{\!
 \begin{array}{ll}
  {P}_k^{q(i)}(T,\mathbf{o}^{q(i)}_k(n)), \quad\qquad\qquad\qquad \textrm{if } q(i)\!=\!q(j),\\
  {P}_k^{q(i)}(T,\mathbf{o}^{q(i)}_k(n)){P}_k^{q(j)}(T,\mathbf{o}^{q(j)}_k(n)), \; \textrm{if } q(i)\!\ne\!q(j),\\
 \end{array}\right.
 \end{array}
\end{eqnarray}
where $\mathbf{o}$ is the concatenation of the observed state vectors of all the clusters. Based on (\ref{eq_ett}) and (\ref{eq_prob_free_link}), we can obtain the 
spectrum-aware link delay metric for $(i,j)_k$ at slot $n$ as follows:
\begin{equation}
 \label{eq_ett_adjusted_channel}
  \begin{array}{ll}
  d_{{(i,j)_k}}(\mathbf{o}(n))\!=\!T\left(1\!-\!{P}_k^{i,j}(\mathbf{o}^{q(i)}_k(n),\mathbf{o}^{q(j)}_k(n))\right)
  +d^{\textrm{ETT}}_{(i,j)_k}{P}_k^{i,j}(\mathbf{o}^{q(i)}_k(n),\mathbf{o}^{q(j)}_k(n)),
   \end{array}
\end{equation}
where $\mathbf{o}(n)$ represents the joint state of the entire secondary network at slot $n$.

Now, we consider the impact of the MAC protocol on the state of link availability. Let $\mathcal{A}_i\!=\!\mathcal{N}_i\backslash\{i\}\!\times\!\mathcal{K}$ denote the set of 
candidate actions for SU $i$. Due to the channel instability, it is difficult to directly adopt MAC protocols based on single-channel random access with exponential backoff 
in the secondary network. Instead, we consider that the contention over each channel is resolved through a reservation mechanism over the common control channel. 
We consider that the negotiating phase over the control channel is divided into $K$ subslots, and the SUs compete for channel $k$ in the corresponding subslot by sending 
Request-To-Send (RTS) packets and listening to Clear-To-Send (CTS) packets from their target relay SUs (Figure \ref{fig1}). Since more than one RTS sent in SU $i$'s 
neighborhood over the same channel will result in collision, the channel negotiation can be considered as a random access mechanism which is similar to slotted-ALOHA.
If channel $k$ is free, the probability of SU $i$ successfully sending the RTS packet over channel $k$ after taking action $a_i$ in $\mathcal{N}_i$ can be written as:
\begin{equation}
 \label{eq_aloha_action_sender}
 P^{k}_i(\mathbf{a}_{\mathcal{N}_i})= I(a_{i,2},k)\prod\limits_{m\in{\mathcal{N}_i\backslash\{i\}}}\left(1-I(a_{m,2},k)\right),
\end{equation}
where $\mathbf{a}_{\mathcal{N}_i}$ is the joint SU action in $\mathcal{N}_i$ and $I(x,y)$ is the indicator function. $I(x,y)\!=\!0$ if $x\!\ne\!y$ and $I(x,y)\!=\!1$ if 
$x\!=\!y$. Similarly, considering the existence of hidden terminals, the probability of SU $j$ successfully receiving the RTS packet from SU $i$ over channel $k$ can be written 
as:
\begin{equation}
 \label{eq_aloha_action_receiver}
 P^{k}_j(\mathbf{a}_{\mathcal{N}_j})= I(a_{i,2},k)\prod\limits_{m\in{\mathcal{N}_j\backslash\{i,j\}}}\left(1-I(a_{m,2},k)\right).
\end{equation}
Based on (\ref{eq_aloha_action_sender}) and (\ref{eq_aloha_action_receiver}), we can express (\ref{eq_ett_adjusted_channel}) under the joint actions $\mathbf{a}_{\mathcal{N}_i}$
and $\mathbf{a}_{\mathcal{N}_j}$ as follows:
\begin{eqnarray}
 \label{eq_link_delay_adjusted}
  \begin{array}{ll}
 d_{(i,j)_k}(\mathbf{o}(n), \mathbf{a}_{\mathcal{N}_i}, \mathbf{a}_{\mathcal{N}_j})=
 T\!\left(1\!-\!P^{k}_i(\mathbf{a}_{\mathcal{N}_i})P^{k}_j(\mathbf{a}_{\mathcal{N}_j})\right)
  +d_{(i,j)_k}(\mathbf{o}^{q(i)}(n),\mathbf{o}^{q(j)}(n))P^{k}_i(\mathbf{a}_{\mathcal{N}_i})P^{k}_j(\mathbf{a}_{\mathcal{N}_j}).
 \end{array}
\end{eqnarray}

In addition to the delay caused by the SU actions following the proposed DSA-MAC, we also need to consider the delay caused by interference between 
multiple flows in the CRN. We assume that each SU can only respond to one randomly chosen RTS during a transmission 
slot. For the proposed DSA-MAC, the number of potential links that can be established to SU $i$ is:
\begin{equation}
 \label{eq_num_rts}
 N_i(\mathbf{a})=\sum\limits_{k\in{\mathcal{K}}}\left(\sum\limits_{m\in{\mathcal{N}_i\backslash\{i\}}}P^{k}_i(\mathbf{a}_{\mathcal{N}_i})
 P^{k}_m(\mathbf{a}_{\mathcal{N}_m})\right),
\end{equation}
where $\mathbf{a}$ is the joint action of all the SUs. According to the queueing delay model based on round-robin packet processing \cite{6599059},
we need to adjust the expected link delay in (\ref{eq_link_delay_adjusted}) by 
substituting $d^{\textrm{ETT}}_{(i,j)_k}$ in (\ref{eq_ett_adjusted_channel}) with $N_i(\mathbf{a})d^{\textrm{ETT}}_{(i,j)_k}$:
  \begin{equation}
  \label{eq_SU_delay} 
    \begin{array}{ll}
  d_i(\mathbf{o}(n), \mathbf{a})\!=\!T\!\Big(1\!-\!P^{k}_i(\mathbf{a}_{\mathcal{N}_i})P^{k}_j(\mathbf{a}_{\mathcal{N}_j})\Big)
  \!+\!\Big(T\Big(1-{P}_k^{i,j}(\mathbf{o}(n))\Big)\!+\!N_i(\mathbf{a})d^{\textrm{ETT}}_{(i,j)_k}{P}_k^{i,j}(\mathbf{o}(n))\Big)
  P^{k}_i(\mathbf{a}_{\mathcal{N}_i})P^{k}_j(\mathbf{a}_{\mathcal{N}_j}),   
   \end{array}
  \end{equation}
\vspace{-10.5pt}
\subsection{Link Quality Metric}
Let $\mathcal{P}(i_0,i_L)=\{(i_0,i_1)_{k_0}, (i_1,i_2)_{k_{1}},\ldots, (i_{L-1},i_{L})_{k_{L-1}}\}$ denote the path formed by a sequence of links between SU $i_0$ and SU $i_L$.
According to Section \ref{sub_sec_node_behavior}, the additional path delay after including SU $i$ into $\mathcal{P}(i_0,i_L)$ is jointly determined by the cluster states of its 
neighbor nodes and the joint action of its two-tier neighbor nodes, see (\ref{eq_link_delay_adjusted}) and (\ref{eq_num_rts}). Based on 
(\ref{eq_ett})-(\ref{eq_num_rts}), we can express the link added by SU $i$ as a function of the joint action of all the SUs $\mathbf{a}=(a_1,\ldots,a_{|\mathcal{N}|})$ in 
(\ref{eq_SU_delay}).
Combining the metrics of the adjusted link delay in (\ref{eq_SU_delay}) and the relay distance advancement in (\ref{eq_distance_advancement}), we can define the instant local utility of SU $i$ as 
a function of the joint state $\mathbf{o}(n)$ and the joint action $\mathbf{a}$ in the secondary network in (\ref{eq_local_util}):
\begin{equation}
 \label{eq_local_util}
  u_i(\mathbf{o}(n),\mathbf{a})=\frac{A_i(a_{i})}{d_{i}(\mathbf{o}(n), \mathbf{a})}.
\end{equation}
According to (\ref{eq_distance_advancement}) and (\ref{eq_SU_delay}), $u_i(\mathbf{o}(n),\mathbf{a})\!\ge\!0$. With (\ref{eq_local_util}), a normal SU $i_0$ measures the quality of its path 
$\mathcal{P}(i_0, i_L)$ as the expected average of the cumulative link utility along the path as follows:
\begin{eqnarray}
 \label{eq_sum_expected_util}
 \begin{array}{ll}
  U_{{\mathcal{P}}(i_0, i_L)}=
  \displaystyle\lim\limits_{\tau\rightarrow\infty}\frac{1}{\tau}E_{\mathbf{o}}\left[\sum\limits_{n=0}^{\tau\!-\!1}\sum_{j{\in}{\mathcal{P}(i_0, i_L)}}u_{j}
  (\mathbf{o}(n),\mathbf{a}(n))\bigg|\mathbf{o}(0)=\mathbf{o}\right].
  \end{array}
\end{eqnarray}
Then for a normal SU, the goal of its relay-selection scheme is to maximize the value of $U_{{\mathcal{P}}(i_0, i_L)}$. 

\vspace{-10.5pt}
\subsection{Impact of Malicious SUs} 
We consider that in the CRN, no SU is superior to the other SUs in obtaining network information. As a result, both the normal SUs and the malicious SUs make their relay-channel 
selection decisions based on the same level of local information. For a malicious relay SU $j$ in path $\mathcal{P}(i_0,i_L)$, the goal is to cause delay as much as 
possible by minimizing the expected cumulative utility $U_{\mathcal{P}(j,i_L)}$ while avoiding being detected as an attacker. To avoid detection, SU $j$ disguises
itself by complying with most of the routing rules in the network layer. Based on its local channel state record, SU $j$ performs the RPU-like attacks by 
violating the interweaving DSA rule and forwarding the packet over the link that has the highest probability of being at state 
\emph{Busy}. SU $j$ may also attempt to send packets to the neighbor SUs which experience larger delay due to channel contention caused by flow intersection. Since 
normal SUs are limited by the number of the equipped radio interfaces, they are not able to passively monitor the neighbors' behaviors.
Also, due to imperfect information about the instantaneous channel states, normal SUs may have difficulties in discerning the delay due to attacks from the delay due to 
PU activities.

Since the SUs cannot exchange the routing information (e.g., local utility and relay-selection decision) with the entire CRN, to form an efficient path they mainly rely on the 
information exchanged between the neighbors. From the perspective of malicious SUs, such a situation of information locality allows them to provide fake information by
distorting the announced value of the expected cumulative link utility for sub-route $\mathcal{P}(j,i_L)$ and induce the normal neighbors to forward packets to them. 
Malicious SUs behave similarly to the SH attackers. Since the operation of information distortion heavily 
depends on the routing scheme adopted by the normal SUs, we will provide more details of this type of attack in the following sections.
\vspace{-10.5pt}
\section{Robust Routing Based on Stochastic Game}
\label{sec_routing_game}
For ease of presentation, in this section we will temporarily ignore the possibility of information distortion by malicious relays and assume truthful information exchange 
between neighbor SUs. Following our discussion of the node behavior and the link quality metric in (\ref{eq_local_util}), we analyze the routing mechanism using a game theoretic model, which 
explicitly addresses the interaction between the normal and the malicious SUs.

\vspace{-10.5pt}
\subsection{Relay Selection as a Stochastic Game}
We define the secondary network global state as the concatenation of the state vectors from all the clusters: $\mathbf{o}=\mathbf{o}^1\|\cdots\|\mathbf{o}^q\|\cdots\|\mathbf{o}^Q$,
where $q=1,\ldots,Q$ is the cluster index. With a slight abuse of notation, we omit the index of the sink SU $i_L$ and denote a path starting from SU $i$ as $\mathcal{P}_i$. 
Then, from Section \ref{sub_sec_dsa}, the evolution of the joint states of any SU sequence retains the Markovian 
property, as stated in Proposition \ref{prop_mdp}:
\vspace{-8.5pt}
\begin{Proposition}
 \label{prop_mdp}
 For any sequence of SUs $\mathcal{P}_{i}$, its joint observed state vector $\mathop{\|}_{q:j\in q(j),\forall j\in\mathcal{P}_i}\mathbf{o}^q$ forms a Markov chain, of 
 which the transition of each state element is independent of the SU actions and can be described by (\ref{eq_sensing_state}).
\end{Proposition}
\vspace{-8.5pt}
The instant utility of path $\mathcal{P}_i$ can be obtained from (\ref{eq_sum_expected_util}) as:
\begin{eqnarray}
 \label{eq_inst_util_path}
  u_{{\mathcal{P}}_i}(\mathbf{o}(n),\mathbf{a}(n))\!=\!u_{i}(\mathbf{o}(n),\mathbf{a}(n))\!+\!\sum_{j{\in}{\mathcal{P}}_i\!\backslash\{i\}}\!u_{j}\!(\mathbf{o}(n),\mathbf{a}(n)).
\end{eqnarray}
For conciseness, we use $U_{\mathcal{P}_i}$ to represent the value of $U_{\mathcal{P}(i, i_L)}$ in (\ref{eq_sum_expected_util}). Since an SU only controls its own decision of 
choosing the next-hop and observes its local utility, the path quality evaluation by SU $i$ will depend on the utility information provided by the next-hop SU. Then, based on 
the path utility in (\ref{eq_inst_util_path}) and Proposition \ref{prop_mdp}, we can define a stochastic routing game in the secondary network as a five-tuple multi-agent 
Markov Decision Process (MDP) \cite{1997Altman}: 
\vspace{-8.5pt}
\begin{Definition}[Stochastic routing game]
 \label{Def_SG}
  The SUs in the CRN form a general-sum stochastic game in the form of a five-tuple: $\mathcal{G}_r\!=\!{\langle}{\mathcal{N}}, {\mathcal{O}}, {\mathcal{A}}, \{u_{\mathcal{P}_i}\}_{i{\in}  
  {\mathcal{N}}}, P(\mathbf{o}'|\mathbf{o}){\rangle}$, in which
\begin{itemize}
 \item ${\mathcal{N}}$ is the set of SUs.
 \item ${\mathcal{O}}$ is the space of the concatenated cluster state vectors.
 \item ${\mathcal{A}}=\times_{i\in{\mathcal{N}}}{{\mathcal{A}}_{i}}$ is the set of joint actions of the SUs.
 \item $u_{\mathcal{P}_i}: {\mathcal{O}}\times{\mathcal{A}}\rightarrow \mathbb{R}$ is the instantaneous utility of the path starting from SU $i$ as in 
 (\ref{eq_inst_util_path}).
 \item $P: {\mathcal{O}}\times{\mathcal{O}}\rightarrow [0,1]$ is the state transition map. 
\end{itemize}
\end{Definition}
\vspace{-8.5pt}
Let $a_{-i}$ denote the joint actions of all SUs except SU $i$, $\pmb\pi_i(\mathbf{o})\!=\!(\pi_i(\mathbf{o},a)\!:\!a\!\in\!{\mathcal{A}}_i)$ denote the 
mixed strategy of SU $i$ at state $\mathbf{o}$, and $\pmb\pi_{-i}(\mathbf{o})\!=\!(\pi_i(\mathbf{o},a_{-i})\!:\!a_{-i}\!\in\!{\mathcal{A}}_{-i})$ denote the mixed strategy of all SUs
except SU $i$ at state $\mathbf{o}$. We note that given the SUs' joint strategy, $\pmb\pi=(\pmb\pi_i(\mathbf{o}),\pmb\pi_{-i}(\mathbf{o}):\mathbf{o}\in\mathcal{O})$, 
the goal of normal SU $i$ is to maximize its expected average utility in (\ref{eq_sum_expected_util}), while the goal of malicious SU $m$ is to minimize the average 
utility. Given $\pmb\pi$, we have:
\begin{eqnarray}
 \label{eq_strategy_util}
 \begin{array}{ll}
 U_{\mathcal{P}_i}(\mathbf{o}, \pmb\pi)=
  \displaystyle\lim\limits_{\tau\rightarrow\infty}\frac{1}{\tau}E_{\mathbf{o},\pmb\pi}\Big\{\!\sum\limits_{n=0}^{\tau\!-\!1}\sum_{j{\in}{\mathcal{P}(i_0, i_L)}}u_{j}
  (\mathbf{o}(n),\mathbf{a}(n))\Big|\mathbf{o}(0)\!=\!\mathbf{o}\!\Big\}.
  \end{array}
\end{eqnarray}
With (\ref{eq_inst_util_path}) and (\ref{eq_strategy_util}), we can define the Nash Equilibrium (NE) of the game as:
\vspace{-8.5pt}
\begin{Definition}[NE] 
 \label{Def_NE}
 $\pmb\pi^*=(\pmb{\pi}^*_i, \pmb{\pi}^*_{-i})$ is an NE for $\mathcal{G}_r$, if $\forall i\in\mathcal{N}$ and $\forall \mathbf{o}{\in}{\mathcal{O}}$ the following 
 conditions are satisfied for any $\pmb\pi_i$:
 \begin{equation}
 \label{eq_ne}
 \left\{
 \begin{array}{ll}
  U_{\mathcal{P}_i}(\mathbf{o}, \pmb\pi^*_i, \pmb\pi^*_{-i})\ge U_{\mathcal{P}_i}(\mathbf{o}, \pmb\pi_i, \pmb\pi^*_{-i}), \textrm{ if SU } i \textrm{ is normal},\\
  U_{\mathcal{P}_i}(\mathbf{o}, \pmb\pi^*_i, \pmb\pi^*_{-i})\le U_{\mathcal{P}_i}(\mathbf{o}, \pmb\pi_i, \pmb\pi^*_{-i}), \textrm{ if SU } i \textrm{ is malicious}.
 \end{array}
 \right.\nonumber
 \end{equation}
\end{Definition}
\vspace{-4.5pt}
Observing (\ref{eq_inst_util_path}), we note that game $\mathcal{G}_r$ differs from a typical stochastic game because the instantaneous individual payoff is determined by not only 
the local link utility, but also the utility of the sub-route starting from the next-hop SU. Therefore, to obtain the NE for $\mathcal{G}_r$, the SUs 
are required to know the sub-route utility of their next-hop nodes. In order to examine the property of the NE for $\mathcal{G}_r$, we introduce the concept of the bias value in 
a multi-agent MDP:
\vspace{-8.5pt}
\begin{Definition}[Bias value]
 \label{Def_BV}
 With initial state $\mathbf{o}$ and policy $\pmb\pi\!=\!(\pmb\pi_i,\pmb\pi_{-i})$, the bias value of SU $i$ is the expected accumulated difference between its 
 instantaneous and stationary utilities: 
  \begin{eqnarray}
  \label{eq_bv}
  \begin{array}{ll}
  h_{\mathcal{P}_i}(\mathbf{o},\pmb\pi)=
  \displaystyle\lim\limits_{\tau\rightarrow\infty}\!E\Big\{\!\sum_{n=0}^{\tau-1}\!\big(u_{\mathcal{P}_i}(\mathbf{o}(n),\pmb\pi)\!-\!
  U_{\mathcal{P}_i}(\mathbf{o}(n),\pmb\pi)\big)\big|\mathbf{o}(0)\!=\!\mathbf{o}\!\Big\}.
  \end{array}
 \end{eqnarray} 
\end{Definition}

Based on (\ref{eq_transition_matrix}) and Proposition \ref{prop_mdp}, we can readily conclude that game $\mathcal{G}_r$ in the sense of a multi-agent MDP is ergodic/recurrent
\cite{puterman2009markov}. Then, using the bias value in Definition \ref{Def_BV}, we introduce the representation of an average utility MDP in the form of
the Bellman optimality equation:
\begin{Lemma}
 \label{lemma1}
 Regardless of the initial state $\mathbf{o}$, the bias value of each SU in game $\mathcal{G}_r$ is constant given any stationary policy $\pmb\pi$ and can be expressed as:
 \begin{eqnarray}
  \label{eq_lemma1}
  h_{\!\mathcal{P}_i}(\mathbf{o},\pmb\pi)\!=\!u_{\!\mathcal{P}_i}(\mathbf{o},\pmb\pi)\!-\!U_{\!\mathcal{P}_i}(\mathbf{o},\pmb\pi)\!+\!\sum_{\mathbf{o}'}P(\mathbf{o}'|
  \mathbf{o})h_{\!\mathcal{P}_i}(\mathbf{o}',\pmb\pi).
 \end{eqnarray}
\end{Lemma}
\begin{proof}
 According to Proposition \ref{prop_mdp}, the transition of the observed states is independent of the SUs' actions. Therefore, according to (\ref{eq_transition_matrix}), for any deterministic strategy 
 $\mathbf{a}$, the underlying Markov chain converges to the same limiting distribution and thus is ergodic. Observing (\ref{eq_local_util}), we note that the instantaneous 
 local link utility $u_i$ is bounded for a finite number of relays. Then, for a sub-route $\mathcal{P}_i$, the expectation and summation in (\ref{eq_strategy_util}) is interchangeable.
 From (\ref{eq_strategy_util}) we obtain:
  \begin{align}
  \label{eq_lemma_prov_1}
  U_{\mathcal{P}_i}(\mathbf{o},\pmb\pi)
  \!=\!\displaystyle\lim\limits_{\tau\rightarrow\infty}\frac{1}{\tau}\sum_{n=0}^{\tau-1}E_{\mathbf{o}}\Big(E_{\pmb\pi}\Big(
  \sum\limits_{j\in\mathcal{P}_i}u_j(\mathbf{o}(n),\mathbf{a}(n))\Big)\Big\vert\mathbf{o}(0)\!=\!\mathbf{o}\Big)   
  =\displaystyle\lim\limits_{\tau\rightarrow\infty}\frac{1}{\tau}\sum_{n=0}^{\tau-1}P(\mathbf{o}'|\mathbf{o})\Big(
  \sum\limits_{j\in\mathcal{P}_i}u_j(\mathbf{o}',\pmb\pi)\Big).
\end{align}
where 
  \begin{eqnarray}
  \label{eq_lemma_prov_2}
  \begin{array}{ll}
  u_{j}(\mathbf{o},\pmb\pi)
  \!=\!\displaystyle\sum_{a_1\in\mathcal{A}_1}\!\cdots\!\sum_{a_{|\mathcal{N}|}\in\mathcal{A}_{|\mathcal{N}|}}\Big(
  u_j(\mathbf{o},\mathbf{a})\times\pmb\pi_1\times\cdots\times\pmb\pi_{|\mathcal{N}|}\Big).
  \end{array}
\end{eqnarray}
 Therefore, with respect to the stationary joint strategy $\pmb\pi$,  each SU's state-value evolution in game $\mathcal{G}_r$
 is reduced to a finite-state, recurrent Markov reward process. Then, Lemma \ref{lemma1} immediately follows Theorem 8.2.6 of \cite{puterman2009markov}.
\end{proof}

By fixing the observed channel state as $\mathbf{o}$ in the stochastic game, we define the stage game of $\mathcal{G}_r$ at state $\mathbf{o}$ as $\mathcal{G}_r(\mathbf{o})=
\langle\mathcal{N},\mathcal{A}, \{u_{\mathcal{P}_i}(\mathbf{o})\}_{i\in\mathcal{N}}\rangle$. $\mathcal{G}_r(\mathbf{o})$ is a normal-form repeated game with normal SUs aiming
at maximizing their instantaneous path utilities and malicious SUs aiming at minimizing the instantaneous path utilities at state $\mathbf{o}$. Based on Lemma \ref{lemma1}, we 
can derive the following results on the NE points of $\mathcal{G}_r$:
\vspace{-8.5pt}
\begin{Theorem}
 \label{thm_BEV}
 (i) $\pmb\pi^*$ is an NE of $\mathcal{G}_r$, only if the following conditions are satisfied $\forall\pmb\pi_i$:
  \begin{eqnarray}
   \label{neq_bev_normal}
   \begin{array}{ll}
   h_{\mathcal{P}_i}(\mathbf{o},\pmb\pi^*)\ge
   u_{\mathcal{P}_i}(\mathbf{o},\pmb\pi_i,\pmb\pi_{-i}^*)\!-\!U_{\mathcal{P}_i}(\mathbf{o},\pmb\pi^*)\!+\!\displaystyle\sum\limits_{\mathbf{o}'}
   P(\mathbf{o}'|\mathbf{o})h_{\mathcal{P}_i}(\mathbf{o}',\pmb\pi^*),
   \end{array}
  \end{eqnarray}
 \vspace{-6mm}
  \begin{eqnarray}
   \label{neq_bev_malicious}
   \begin{array}{ll}
   h_{\mathcal{P}_j}(\mathbf{o},\pmb\pi^*)\le
   u_{\mathcal{P}_j}(\mathbf{o},\pmb\pi_j,\pmb\pi_{\!-\!j}^*)\!-\!U_{\!\mathcal{P}_j}(\mathbf{o},\pmb\pi^*)\!+\!\displaystyle\sum\limits_{\mathbf{o}'}
   P(\mathbf{o}'|\mathbf{o})h_{\mathcal{P}_m}(\mathbf{o}',\pmb\pi^*),
   \end{array}
  \end{eqnarray}
for every normal SU $i$ and malicious SU $j$.

(ii) $\pi^*(\mathbf{o})$ is also an NE strategy of $\mathcal{G}_r(\mathbf{o})$. The NE strategies of all the stage games, $\mathcal{G}_r(\mathbf{o}\!:\!\forall 
\mathbf{o}\!\in\!\mathcal{O})$, constitute an NE strategy of $\mathcal{G}_r$.
\end{Theorem}
\vspace{-8.5pt}
\begin{proof}
 \label{pro_BEV}
 See Appendix \ref{app_thm_BEV}.
\end{proof}
\vspace{-8.5pt}
\begin{Remark} 
\label{remark_thm_bev}
\rm
Theorem \ref{thm_BEV} establishes the equivalence between the NE strategies of $\mathcal{G}_r$ and the group of NE strategies of its corresponding stage games. It is worth 
noting that Theorem \ref{thm_BEV} is based on Proposition \ref{prop_mdp}. In this case, all the NE of the stochastic game are the stationary Markov perfect equilibria. In 
contrast, for a general-case stochastic game where the state transition is usually a function of the players' actions, the equality in (\ref{eq_lemma1}) 
may not hold except for the equilibrium strategies, and the second property in Theorem \ref{thm_BEV} does not exist.
\qed
\end{Remark}
\vspace{-8.5pt}
Due to the overhead caused by information flooding, it is unrealistic for the SUs to frequently exchange the information about their private actions and utilities with the SUs beyond 
the one-hop neighbors. To determine the level of information exchange in the routing game, we consider another multi-agent MDP based on  each SU's local utility $u_i$. We 
define the MDP as $\mathcal{G}_l=\langle\mathcal{N}, \mathcal{O}, \mathcal{A}, \{u_{i}\}_{i\in\mathcal{N}}, P(\mathbf{o}'|\mathbf{o})\rangle$. Let $h_i(\mathbf{o},\pmb\pi)$ 
denote the bias value of $\mathcal{G}_l$ and $U_i(\mathbf{o},\pmb\pi)$ denote the average gain value of $\mathcal{G}_l$. Then, we can show that Lemma \ref{lemma1} also 
applies to the pair of $h_i$ and $U_i$ in $\mathcal{G}_l$. Let $\pmb\pi_{i,1}$ denote the strategy of SU $i$ for selecting the next hop, $\pmb\pi_{i,2}$ denote the  strategy of 
SU $i$ for selecting the transmitting channel, and $\mathcal{P}_{a_{i,1}}$ denote the sub-route in $\mathcal{P}_{i}$ starting from the node that is chosen by SU $i$ with action 
$a_{i,1}$. Based on Lemma \ref{lemma1} and Theorem \ref{thm_BEV}, we can show that $\mathcal{G}_r$ can be decomposed into a layered multi-agent MDP in the 
following theorem:
\vspace{-8.5pt}
\begin{Theorem}
 \label{thm_layered_mdp}
 (i) With stationary joint policy $\pmb\pi$, the relay selection process of SU $i$ can be expressed as (\ref{eq_layered_mdp}):
\begin{equation}
 \label{eq_layered_mdp}
 \begin{array}{ll}  
  h_{\mathcal{P}_i}(\mathbf{o},\pmb\pi)\!+\!U_{\mathcal{P}_i}(\mathbf{o},\pmb\pi)
  =\\
  u_i(\mathbf{o},\pmb\pi)\!+\!\sum\limits_{\mathbf{o}'}\!P(\mathbf{o}'|\mathbf{o})h_{i}(\mathbf{o}',\pmb\pi)
  \!+\!E_{\pmb\pi_{i,1}}\Big\{u_{\mathcal{P}_{a_{i,1}}}
  (\mathbf{o},a_{i,1}, \pmb\pi_{i,2}, \pmb\pi_{-i})
  \!+\!\sum\limits_{\mathbf{o}'}P(\mathbf{o}'|\mathbf{o})h_{\mathcal{P}_{a_{i,1}}}
 (\mathbf{o},a_{i,1}, \pmb\pi_{i,2}, \pmb\pi_{-i})\Big\}.
 \end{array}
\end{equation}

  (ii) Strategy $\tilde{\pmb\pi}$ is an NE point of $\mathcal{G}_r$ when for any normal SU $i$ and malicious SU $j$,
  \begin{equation}
   \label{eq_NE_equivalence}
   \left\{\!
   \begin{array}{ll}
   \tilde{\pmb\pi}_i\!=\!\arg\max\limits_{{\pmb\pi}_i}(h_{\mathcal{P}_i}(\mathbf{o},{\pmb\pi}_i,\tilde{\pmb\pi}_{-i})\!+\!U_{\mathcal{P}_i}(\mathbf{o},{\pmb\pi}_i,\tilde{\pmb\pi}_{-i})),\\
   \tilde{\pmb\pi}_j\!=\!\arg\min\limits_{{\pmb\pi}_j}(h_{\mathcal{P}_j}(\mathbf{o},{\pmb\pi}_j,\tilde{\pmb\pi}_{-j})\!+\!U_{\mathcal{P}_j}(\mathbf{o},{\pmb\pi}_j,\tilde{\pmb\pi}_{-j})),
   \end{array}\right.
  \end{equation}
\end{Theorem}
\vspace{-8.5pt}
\begin{proof}
 \label{pro_layered_mdp}
 See Appendix \ref{app_thm_layered_mdp}.
\end{proof}
\vspace{-8.5pt}
\begin{Remark} 
\label{remark_layered_mdp}
\rm
Theorem \ref{thm_layered_mdp} shows that given a stationary joint policy $\pmb\pi$, the relay-selecting process of an SU is composed of two value iteration processes in the form 
of the Bellman optimality equation.  The first one is determined by the local multi-agent MDP $\mathcal{G}_l$, and the second one is determined by the sub-path starting 
from the selected next-hop SU $a_{i,1}$ 
in $\mathcal{G}_r$. Furthermore, the second Bellman optimality equation for SU $a_{i,1}$ can be decomposed into the same two-layer form as (\ref{eq_layered_mdp})
with respect to its own decision on next-hop selection.
\qed
\end{Remark}
\vspace{-8.5pt}
According to Theorem \ref{thm_layered_mdp}, to derive its local NE strategy $\pmb\pi^*_i$, SU
$i$ needs its neighbor nodes $j\!\in\!\mathcal{N}_i\backslash\{i\}$ to truthfully provide the information on the equilibrium value of $h_{\mathcal{P}_j}(\mathbf{o},a_{i,1}\!=\!j, 
\pmb\pi_{i,2}, \pmb\pi^*_{-i})\!+\!U_{\mathcal{P}_j}(\mathbf{o},a_{i,1}\!=\!j, \pmb\pi_{i,2}, \pmb\pi^*_{-i})$. It requires that the
NE for game $\mathcal{G}_r$ is solved through backward induction. Observing (\ref{eq_strategy_util}) and (\ref{eq_bv}), it is straightforward to show that when stochastic game 
$\mathcal{G}_r$ is reduced to a stage game $\mathcal{G}_r(\mathbf{o})$ with a single state $\mathbf{o}$, providing the value $h_{\mathcal{P}_j}(\mathbf{o},
a_{i,1}\!=\!j, \pmb\pi_{i,2}, \pmb\pi^*_{-i})\!+\!U_{\mathcal{P}_j}(\mathbf{o},a_{i,1}\!=\!j, \pmb\pi_{i,2}, \pmb\pi^*_{-i})$ is equivalent to providing the  
value $u_{\mathcal{P}_j}(\mathbf{o},a_{i,1}\!=\!j, \pmb\pi_{i,2}, \pmb\pi^*_{-i})$. Such an observation paves the way for developing a strategy-learning method based on limited
information exchange between the SUs.

\vspace{-10.5pt}
\subsection{Strategy Learning with Truthful Information Exchange}
\label{sec_learning_truthful_game}
According to Theorem \ref{thm_BEV}, an NE for the stochastic routing game can be constructed based on the state-dependent NE strategies for each stage routing game with fixed
estimated channel states.
Therefore, we consider a stage routing game at state $\mathbf{o}$: $\mathcal{G}_r(\mathbf{o})\!=\!\langle\mathcal{N},\mathcal{A}, \{u_{\mathcal{P}_i}
(\mathbf{o})\}_{i\in\mathcal{N}}\rangle$, where $u_{\mathcal{P}_i}(\mathbf{o},\mathbf{a})\!=\!u_{i}(\mathbf{o},\mathbf{a})\!+\!u_{\mathcal{P}_{a_{i,1}}}(\mathbf{o},\mathbf{a})$. Based
on Theorem \ref{thm_layered_mdp}, the NE for $\mathcal{G}_r(\mathbf{o})$ is achieved when each normal SU $i$ and malicious SU $j$ play the strategies $\pmb\pi^*$ that satisfy the following 
conditions:
\begin{equation}
 \label{eq_NE_stage}
 \left\{
 \begin{array}{ll}
 U^*_{\mathcal{P}_i}=\max\limits_{\pmb\pi_i}\Big(u_{i}(\mathbf{o},\pmb\pi_i,\pmb\pi^*_{-i})
 +E_{\pmb\pi_{i,1}}\left\{u_{\mathcal{P}_{a_{i,1}}}(\mathbf{o},a_{i,1}, \pmb\pi_{i,2},\pmb\pi^*_{-i})\right\}\Big),\\
 U^*_{\mathcal{P}_j} =\min\limits_{\pmb\pi_j}\Big(u_{j}(\mathbf{o},\pmb\pi_m,\pmb\pi^*_{-j})
 +E_{\pmb\pi_{j,1}}\left\{u_{\mathcal{P}_{a_{j,1}}}(\mathbf{o},a_{j,1}, \pmb\pi_{j,2},\pmb\pi^*_{-j})\right\}\Big).
 \end{array}
  \right.
\end{equation}

To avoid information flooding, we assume that the SUs do not share with their neighbors the local action information. An SU is only able to share its value of actions by 
exchanging routing request (RREQ) and routing response (RREP) packets with its neighbors.  In this section, we consider that malicious SUs do not provide distorted information. 
Since an SU cannot observe other nodes' actions, we resort to reinforcement learning to obtain the NE under the 
condition of incomplete information. We assume that a stationary joint strategy $\pmb\pi$ is adopted by the SUs in game $\mathcal{G}_r(\mathbf{o})$. Then, we
consider the following action-value learning process for SU $i$:
\begin{eqnarray}
 \label{eq_action_value}
 \begin{array}{ll}
 \tilde{u}^{n+1}_{\mathcal{P}_i}(\mathbf{o}, \mathbf{a}_i)=\tilde{u}^{n}_{\mathcal{P}_i}(\mathbf{o}, \mathbf{a}_i)+
 \alpha(n)I(\mathbf{a}_i(n),\mathbf{a}_i)\left(u_{\mathcal{P}_i}(\mathbf{o}, \mathbf{a}_i(n),\mathbf{a}_{-i}(n))\!-\!\tilde{u}^{n}_{\mathcal{P}_i}(\mathbf{o}, \mathbf{a}_i)\right),
 \end{array}
\end{eqnarray}
where $\tilde{u}^{n}_{\mathcal{P}_i}(\mathbf{o}, \mathbf{a}_i)$ is the expected path utility learned for action $\mathbf{a}_i$ at slot $n$, and $0\!<\!\alpha(n)\!<\!1$ is a 
sequence of learning rates. According to reinforcement learning theory \cite{leslie2003convergent}, if $u_{\mathcal{P}_i}(\mathbf{o}, \mathbf{a}_i(n),\mathbf{a}_{-i}(n))$ is 
perfectly known by SU $i$, $\tilde{u}^{n}_{\mathcal{P}_i}(\mathbf{o}, \mathbf{a}_i)$ converges almost surely to the real value of $u_{\mathcal{P}_i}(\mathbf{o},\mathbf{a}_i,
\pmb\pi_{-i})$, given that all the possible action combinations are visited infinitely often by the SUs and $\alpha(n)$ satisfies the conditions $\sum_{n}\alpha(n)=\infty$ and 
$\sum_{n}\alpha^2(n)<\infty$.

We first assume that an SU $i$ is able to timely calculate the instantaneous accumulated utility of path $\mathcal{P}_i$ based on its local observation of $u_i(\mathbf{o}(n), 
\mathbf{a}(n))$ and the instantaneous sub-path utility $u_{\mathcal{P}_{\mathbf{a}_{i,1}}}(\mathbf{o}(n),\mathbf{a}(n))$, which is fed back by its next-hop SU 
$j=\mathbf{a}_{i,1}$. In this case, (\ref{eq_action_value}) can be adopted by each SU to estimate their action value in stage game $\mathcal{G}_r(\mathbf{o})$. Then, we can adopt
the algorithm of Stochastic Fictitious Play (SFP) \cite{ECTA:ECTA376} for SU $i$ to learn ${\pmb\pi}_i(\mathbf{o}, \mathbf{a}_i,\pmb\pi_{-i})$:
\begin{equation}
 \label{eq_strategy_learning}
 \begin{array}{ll}
 \tilde{\pmb\pi}^{n\!+\!1}_i(\mathbf{o},\mathbf{a}_i)\!=\!\tilde{\pmb\pi}^{n}_i(\mathbf{o},\mathbf{a}_i)\!+\!
 \beta(n)\!\left(\textrm{BR}\!\left(\tilde{\mathbf{u}}^{n}_{\mathcal{P}_i}(\mathbf{o}),\mathbf{a}_i\right)\!-\!\tilde{\pmb\pi}^{n}_i(\mathbf{o},
 \mathbf{a}_i)\right),\\
 \end{array}
\end{equation}
where $\tilde{\mathbf{u}}^n_{\mathcal{P}_i}(\mathbf{o})$ is the vector of utility $\tilde{u}^{n}_{\mathcal{P}_i}(\mathbf{o}, \mathbf{a}_i)$ for all action $\mathbf{a}_i$ at time 
slot $n$, and $\textrm{BR}(\cdot)$ is the perturbed best response strategy of SU $i$ for action $\mathbf{a}_i$ in the form of the Logit function:
\begin{eqnarray}
 \label{eq_perturbed_br}
 \begin{array}{ll}
 \textrm{BR}\left(\tilde{\mathbf{u}}^n_{\mathcal{P}_i}(\mathbf{o}), \mathbf{a}_i\right)=
 \left\{
 \begin{array}{ll}
 \displaystyle\frac{\exp\left(\lambda_i\left(\tilde{{u}}^{n}_{\mathcal{P}_{i}}(\mathbf{o},\mathbf{a}_i)\right)\right)}
 {\sum\nolimits_{\mathbf{b}\in\mathcal{A}_i}\exp\left(\lambda_i\left(\tilde{{u}}^{n}_{\mathcal{P}_{i}}(\mathbf{o},\mathbf{b})\right)\right)}, 
 \;\quad i \textrm{ is normal},\\
 \displaystyle\frac{\exp\left(\lambda_i\left(\tilde{{u}}^{n}_{\mathcal{P}_{i}}(\mathbf{o},\mathbf{a}_i)\right)^{-1}\right)}
 {\sum\nolimits_{\mathbf{b}\in\mathcal{A}_i}\exp\left(\lambda_i\left(\tilde{{u}}^{n}_{\mathcal{P}_{i}}(\mathbf{o},\mathbf{b})\right)^{-1}\right)}, 
 i \textrm{ is malicious}.
 \end{array}\right.
  \end{array}
\end{eqnarray}
The utility learning process in (\ref{eq_action_value}) and the SPF-based strategy learning process in (\ref{eq_strategy_learning}) and (\ref{eq_perturbed_br}) form a two 
timescale learning scheme, which has the following convergence property:
\vspace{-8.5pt}
\begin{Theorem}
 \label{thm_pseudotrajectory}
 If $u_{\mathcal{P}_i}(\mathbf{o},\mathbf{a}(n))$ is known to each SU at every time slot, and the following conditions are satisfied:
 $\lim\limits_{n\rightarrow\infty}\sum_n\alpha(n)\!=\!\infty$, $\lim\limits_{n\rightarrow\infty}\sum_n\alpha^2(n)  \!<\!\infty$, $\lim\limits_{n\rightarrow\infty}\sum_n\beta(n)\!
 =\!\infty$, $\lim\limits_{n\rightarrow\infty}\sum_n\beta^2(n)\!<\!\infty$ and $\lim\limits_{n\rightarrow\infty}(\beta(n)/\alpha(n))\!=\!0$, then 
 $\{\tilde{\pmb\pi}^{n}_i(\mathbf{o},\mathbf{a}_i)\}$ given by the learning process (\ref{eq_action_value})-(\ref{eq_perturbed_br}) converges almost surely to an NE for 
 stage game $\mathcal{G}_r(\mathbf{o})$.
\end{Theorem}
\vspace{-8.5pt}
\begin{proof}
 See Appendix \ref{app_thm_Pseudotraj}.
\end{proof}

Although the learning scheme given by (\ref{eq_strategy_learning}) and (\ref{eq_perturbed_br}) possesses good convergence property, the assumption of perfectly knowing the 
instantaneous path utility is fairly strict. It requires a large amount of signaling to be performed within a single time slot. To address such a problem,
we relax the requirement on information exchange by assuming that SU $i$ only shares its locally estimated value of $u_{\mathcal{P}_i}(\mathbf{o}, \pmb\pi)$ with the 
neighbors. Based on the discussion of (\ref{eq_action_value}), it is obvious that an SU can learn its expected local link utility $u_i(\mathbf{o},\mathbf{a}_i,\pmb\pi_{-i})$ 
through an iteration which is similar to (\ref{eq_action_value}), as long as all possible joint actions are visited infinitely often:
\begin{eqnarray}
 \label{eq_action_value_local}
 \begin{array}{ll}
 \tilde{u}^{n+1}_{i}(\mathbf{o}, \mathbf{a}_i)=
 \tilde{u}^{n}_{i}(\mathbf{o},\mathbf{a}_i)\!+\!
 \alpha(n){I(\mathbf{a}_i(n),\mathbf{a}_i)}\left(u_{i}(\mathbf{o},\mathbf{a}(n))\!-\!\tilde{u}^{n}_{i}(\mathbf{o},\mathbf{a}_i)\right).
 \end{array}
\end{eqnarray}
Using the value of $\tilde{u}^{n}_{i}(\mathbf{o},\mathbf{a}_i)$ and the value of $\tilde{u}^{n}_{\mathcal{P}_{\mathbf{a}_{i,1}}}(\mathbf{o})$ provided by the next-hop SU 
$j=\mathbf{a}_{i,1}$, we introduce the learning scheme for $u_{\mathcal{P}_i}(\mathbf{o}, \pmb\pi)$:
\begin{eqnarray}
 \label{eq_action_value_path}
 \begin{array}{ll}
  \tilde{u}^{n+1}_{\mathcal{P}_i}(\mathbf{o})=\tilde{u}^{n}_{\mathcal{P}_i}(\mathbf{o})+
  \gamma_i(n)\Big(\displaystyle\sum_{\mathbf{a}_i}\tilde{\pmb\pi}^{n}_i(\mathbf{o},\mathbf{a}_i)
  \big(\tilde{u}^{n}_{i}(\mathbf{o},\mathbf{a}_i)+\tilde{u}^{n}_{\mathcal{P}_{\mathbf{a}_{i,1}}}(\mathbf{o})\big)\!-\!\tilde{u}^{n}_{\mathcal{P}_i}(\mathbf{o})\Big).
 \end{array}
\end{eqnarray}
We also modify the learning scheme of (\ref{eq_strategy_learning}) and obtain:
\begin{eqnarray}
 \label{eq_strategy_learning_modified}
 \begin{array}{ll}
 \tilde{\pmb\pi}^{n+1}_i(\mathbf{o},\mathbf{a}_i)\!=\!\tilde{\pmb\pi}^{n}_i(\mathbf{o},\mathbf{a}_i)+
 \beta(n)\left(\textrm{BR}\big(\tilde{\mathbf{u}}^n_{i}(\mathbf{o}, \mathbf{a}_i)+\tilde{\mathbf{u}}^{n}_{\mathcal{P}_{a_{i,1}}}(\mathbf{o}),\mathbf{a}_i\big)-\tilde{\pmb\pi}^{n}_i(\mathbf{o},
 \mathbf{a}_i)\right),\\
 \end{array}
\end{eqnarray}
where $\textrm{BR}(\cdot)$ is the modified perturbed best response strategy:
\begin{equation}
 \label{eq_perturbed_br_modified}
 \begin{array}{ll}
 \textrm{BR}\Big(\tilde{\mathbf{u}}^n_{i}(\mathbf{o})+\tilde{{u}}^{n}_{\mathcal{P}_{a_{i,1}}}(\mathbf{o}), \mathbf{a}_i\Big)\!=\!
 \left\{\!
 \begin{array}{ll}
 \displaystyle\frac{\exp\left(\lambda_i\big(\tilde{u}_i(\mathbf{o}, \mathbf{a}_i)+\tilde{{u}}^{n}_{\mathcal{P}_{a_{i,1}}}(\mathbf{o})\big)\right)}
 {\sum_{\mathbf{b}\in\mathcal{A}_i}\!\exp\left(\lambda_i\big(\tilde{u}_i(\mathbf{o}, \mathbf{b})+\tilde{{u}}^{n}_{\mathcal{P}_{a_{i,1}}}(\mathbf{o})\big)\right)}, 
 \;\;\; i \textrm{ is normal},\\
 \displaystyle\frac{\exp\left(\lambda_i\big(\tilde{u}_i(\mathbf{o}, \mathbf{a}_i)+\tilde{{u}}^{n}_{\mathcal{P}_{a_{i,1}}}(\mathbf{o})\big)^{-1}\right)}
 {\sum_{\mathbf{b}\in\mathcal{A}_i}\!\exp\left(\lambda_i\big(\tilde{u}_i(\mathbf{o}, \mathbf{b})+\tilde{{u}}^{n}_{\mathcal{P}_{a_{i,1}}}(\mathbf{o})\big)^{-1}\right)}, 
 i \textrm{ is malicious}.
 \end{array}\right.
  \end{array}
\end{equation}

The learning scheme defined by (\ref{eq_action_value_local})-(\ref{eq_perturbed_br_modified}) does not require SU $i$ to immediately report the instantaneous path utility to its 
previous-hop SU. However, comparing (\ref{eq_action_value_path}) with (\ref{eq_NE_stage}), we note that (\ref{eq_action_value_path}) provides a biased estimation of 
$u_{\mathcal{P}_i}(\mathbf{o}, \pmb\pi)$, given that $\tilde{\pmb\pi}^{n}_i$ converges.
Since the learned path utility in (\ref{eq_action_value_path}) is a biased estimation, the new learning scheme can only obtain an approximation of the NE 
point of the stage game. The convergence condition of the learning scheme given by (\ref{eq_action_value_local})-(\ref{eq_perturbed_br_modified}) is provided in Theorem 
\ref{thm_convergence_approximation}.
\vspace{-8.5pt}
\begin{Theorem}
 \label{thm_convergence_approximation}
 Assume that the following are satisfied:
 $\lim\limits_{n\rightarrow\infty}\sum\limits_n\alpha(n)\!=\!\infty$, $\lim\limits_{n\rightarrow\infty}\sum\limits_n\alpha^2(n)\!<\!\infty$, 
 $\lim\limits_{n\rightarrow\infty}\sum\limits_n\gamma_i(n)\!=\!\infty$, $\lim\limits_{n\rightarrow\infty}\sum\limits_n\gamma_i^2(n)\!<\!\infty$,
 $\lim\limits_{n\rightarrow\infty}\sum\limits_n\beta(n)\!=\!\infty$, $\lim\limits_{n\rightarrow\infty}\sum\limits_n\beta^2(n)\!<\!\infty$, 
 $\lim\limits_{n\rightarrow\infty}(\gamma_i(n)/\alpha(n))\!=\!0$, $\lim\limits_{n\rightarrow\infty}(\beta(n)/\gamma_i(n))\!=\!0$, 
 and $\lim\limits_{n\rightarrow\infty}(\gamma_i(n)/\gamma_j(n))\!=\!0$, if SU $i$ is closer to the sink SU than 
 SU $j$ in terms of distance. Then, 
 $\{\tilde{\pmb\pi}^{n}_i(\mathbf{o},\mathbf{a}_i)\}$ obtained through the learning process defined by (\ref{eq_action_value_local})-(\ref{eq_perturbed_br_modified}) 
 converges almost surely.
\end{Theorem}
\begin{proof}
 See Appendix \ref{app_thm_convergence_approximation}.
\end{proof}

\vspace{-8.5pt}
\subsection{Strategy Learning with Truth-telling Enforcement}
\label{sec_learning_sophisticated_game}
Now, we consider the situation when the malicious SUs also perform the SH attacks. In this case, the malicious SUs may distort the value of 
$\tilde{u}_{\mathcal{P}_i}(\mathbf{o})$ and report an estimated utility which is much larger than the real value to their neighbor SUs. As a result, a normal neighbor with 
strategy learning scheme in (\ref{eq_strategy_learning_modified}) will choose the malicious SU as its 
relay with a higher probability. Then, the malicious SU will induce the neighbor SUs to forward more packets to them.
To address this situation, we introduce a feedback mechanism for a relay SU to measure the real delay of the path that it chooses toward a sink SU. We consider that 
a normal relay SU $i$ is able to insert a Request-ACK packet into the flows that it serves in random time intervals. SU $i$ records the time stamp for sending the Request-ACK packet. 
When receiving the Request-ACK packet, the corresponding sink SU $L$ replies the Response packet to SU $i$ by including the time 
stamp for reception in the packet. We assume that the data in the Response packet is protected by a pair of keys and is always reliable. With the two time stamps, SU $i$ is able to 
calculate the total delay time of the Request-ACK packet over the sub-path through the next-hop SU $j=\mathbf{a}_{i,1}$ that it chooses with action $\mathbf{a}_{i}$. Let $c_i(\mathbf{a}_{i})$ denote such a delay 
measured by SU $i$. Then, SU $i$ needs to evaluate the trustworthiness of its next-hop SU $j$ based on sequence $\{c^{\hat{n}}_i(\mathbf{a}_{i}(\hat{n}))\}$, in which $\hat{n}$ is a 
time slot for SU $i$ to send a Request-ACK packet.

We consider that with a certain termination condition, the learning scheme given in (\ref{eq_action_value_local})-(\ref{eq_perturbed_br_modified}) can always reach a stationary policy 
$\pmb\pi$. Meanwhile, a malicious SU $m$ shares a fixed distorted value of $\tilde{u}_{\mathcal{P}_m}(\mathbf{o})$ with the neighbors. From the perspective of a normal SU $i$, 
when sending a Request-ACK packet, its relay selection can be considered as a Multi-Arm Bandit (MAB) \cite{ASMB:ASMB874} process, since at slot $\hat{n}$ SU $i$ can 
choose only one neighbor as its relay according to $\mathbf{a}_{i,1}(\hat{n})$, and only the real path delay through relay node $\mathbf{a}_{i,1}(\hat{n})$ can be 
confirmed. We note that the real path delay (i.e., the cost of each arm) is a stochastic function determined by stationary distribution $\pmb\pi$, while the arm selection 
sequence is generated by the local strategy $\pmb\pi_i$. Formally, we can define the MAB for trustworthiness evaluation as follows:
\vspace{-8.5pt}
\begin{Definition}
\label{def_MAB}
For each normal SU $i$, the MAB for trustworthiness evaluation in state $\mathbf{o}$ can be defined by a 4-tuple: $\mathcal{B}_i=\langle\mathcal{A}_i, 
\{c^{\hat{n}}_i(\mathbf{o}, \mathbf{a}_{i}(\hat{n}))\}_{\hat{n}}, \{x(\hat{n})\}_{\hat{n}}, \{\hat{n}\}\rangle$, in which
\begin{itemize}
 \item $\mathcal{A}_i$ is the set of the single-bandit processes and corresponds to the set of actions of SU $i$.
 \item $\{c^{\hat{n}}_i(\mathbf{o}, \mathbf{a}_{i}(\hat{n}))\}_{\hat{n}}$ is the sequence of cost.
 \item $\{x(\hat{n})\!=\!\mathbf{a}_i(\hat{n})\}_{\hat{n}}$ is the sequence of relay (i.e., arm) selection decision.
\end{itemize}
\end{Definition}
It is worth noting that the MAB given in Definition \ref{def_MAB} differs from a typical MAB in that the sequence of arm selection $\{x(\hat{n})\}_{\hat{n}}$ is generated 
following a given policy $\pmb\pi_i$. Therefore, we can consider the MAB process up to time slot $n$ as a utility exploration phase with a given sampling distribution $\pmb\pi_i$.
With the MAB given in Definition \ref{def_MAB}, SU $i$ is able to calculate the accumulated delay for the Request-ACK packets that it sends with action 
$\mathbf{a}_i(n)=\mathbf{a}$:
\begin{equation}
 \label{eq_discounted_delay}
  C^n_i(\mathbf{o}, \mathbf{a})=\left\{
  \begin{array}{ll}
  c^{{n}}_i(\mathbf{o}, \mathbf{a}_{i}({n}))+C^{n-1}_i(\mathbf{o}, \mathbf{a}) \textrm{ if } n\in\{\hat{n}\},\mathbf{a}_{i}({n})=\mathbf{a}\\
  C^{n-1}_i(\mathbf{o}, \mathbf{a}) \qquad\qquad\quad\quad\textrm{ otherwise} ,\\
  \end{array}
  \right.
\end{equation} 
and the sampled frequency of each action is:
\begin{equation}
 \label{eq_sample_frequency}
 Z^n(\mathbf{o}, \mathbf{a})=\frac{1}{n}(I(\mathbf{a}_i(n),\mathbf{a})+(n-1)Z^{n-1}(\mathbf{o}, \mathbf{a})).
\end{equation}
With (\ref{eq_discounted_delay}) and (\ref{eq_sample_frequency}), we can obtain the sampled average path delay of SU $i$ at action $\mathbf{a}(n)\!=\!\mathbf{a}$ as 
$R^n_i(\mathbf{o}, \mathbf{a})\!=\!C^n_i(\mathbf{o}, \mathbf{a})/Z^n_i(\mathbf{o}, \mathbf{a})$. Then
according to \cite{ASMB:ASMB874}, a greedy, sub-optimal mixed strategy for arm allocation to minimize the average path delay can be obtained using the Logit function:
\begin{equation}
  \label{eq_sigma}
 \tilde\sigma^n_i(\mathbf{o}, \mathbf{a})=\frac{\exp\left(\lambda_i (R^n_i\left(\mathbf{o}, \mathbf{a})\right)^{-1}\right)}{\sum_{\mathbf{b}\in\mathcal{A}_i}\exp\left(\lambda_i\left(
 R^n_i(\mathbf{o}, \mathbf{b})\right)^ {-1}\right)},
\end{equation}
which is in a similar form to (\ref{eq_perturbed_br}). $\tilde\sigma^n_i(\mathbf{a})$ does not have to be consistent with the learned equilibrium policy when every SU is honest. 
However, it can represent the ranking value of the trustworthiness of the relay associated with action $\mathbf{a}$. During the estimation of the perturbed best response, 
a normal SU will consider the contribution of the reported sub-path utility by its neighbors in proportion to the trustworthiness credit that it assigns to each neighbor. 
According to (\ref{eq_SFP_1}), a normal SU $i$ modifies its smooth best response objective as follows:
\begin{equation}
 \label{eq_SFP_1_weighted}
 \begin{array}{ll}
 \overline{\textrm{BR}}({\pmb\pi_{-i}})\!=\!
 \arg\max\limits_{\pmb\pi_i}\Big(\sum\limits_{\mathbf{a}_i}\pmb\pi_i(\mathbf{a}_i)\Big({u}_i(\mathbf{o},\pmb\pi(\mathbf{a}_i),\pmb\pi_{-i})\!+\!
 \tilde{\sigma}^n_i(\mathbf{o},\mathbf{a}_i){{u}}_{\mathcal{P}_{a_{i,1}}} (\mathbf{o},\pmb\pi)\Big)
 \!-\!\lambda_i\displaystyle\sum\limits_{\mathbf{a}_i}\pmb\pi_i(\mathbf{a}_i)\log\pmb\pi_i(\mathbf{a}_i)\Big).
 \end{array}
\end{equation}
Then, its perturbed best response strategy can be adjusted as:
\begin{equation}
 \label{eq_perturbed_br_adjusted}
 \begin{array}{ll}
 \textrm{BR}\Big(\tilde{\mathbf{u}}^n_{i}(\mathbf{o})+\tilde{{u}}^{n}_{\mathcal{P}_{a_{i,1}}}(\mathbf{o}), \mathbf{a}_i\Big)=
 \displaystyle\frac{\exp\Big(\lambda_i\Big(\tilde{u}_i(\mathbf{o}, \mathbf{a}_i)+\tilde{\sigma}^n_i(\mathbf{o},\mathbf{a}_i)\tilde{{u}}^{n}_{\mathcal{P}_{a_{i,1}}}(\mathbf{o})\Big)
 \Big)} {\sum_{\mathbf{b}\in\mathcal{A}_i}\exp\Big(\lambda_i\Big(\tilde{u}_i(\mathbf{o}, \mathbf{b})+\tilde{\sigma}^n_i(\mathbf{o},\mathbf{b})
 \tilde{{u}}^{n}_{\mathcal{P}_{a_{i,1}}}(\mathbf{o})\Big)\Big)}, 
  \end{array}
\end{equation}

\section{Simulation Results}
\label{sec_simulation}
\begin{figure}
 \begin{minipage}{.49\textwidth}
  \centering
  \includegraphics[width=0.8\textwidth]{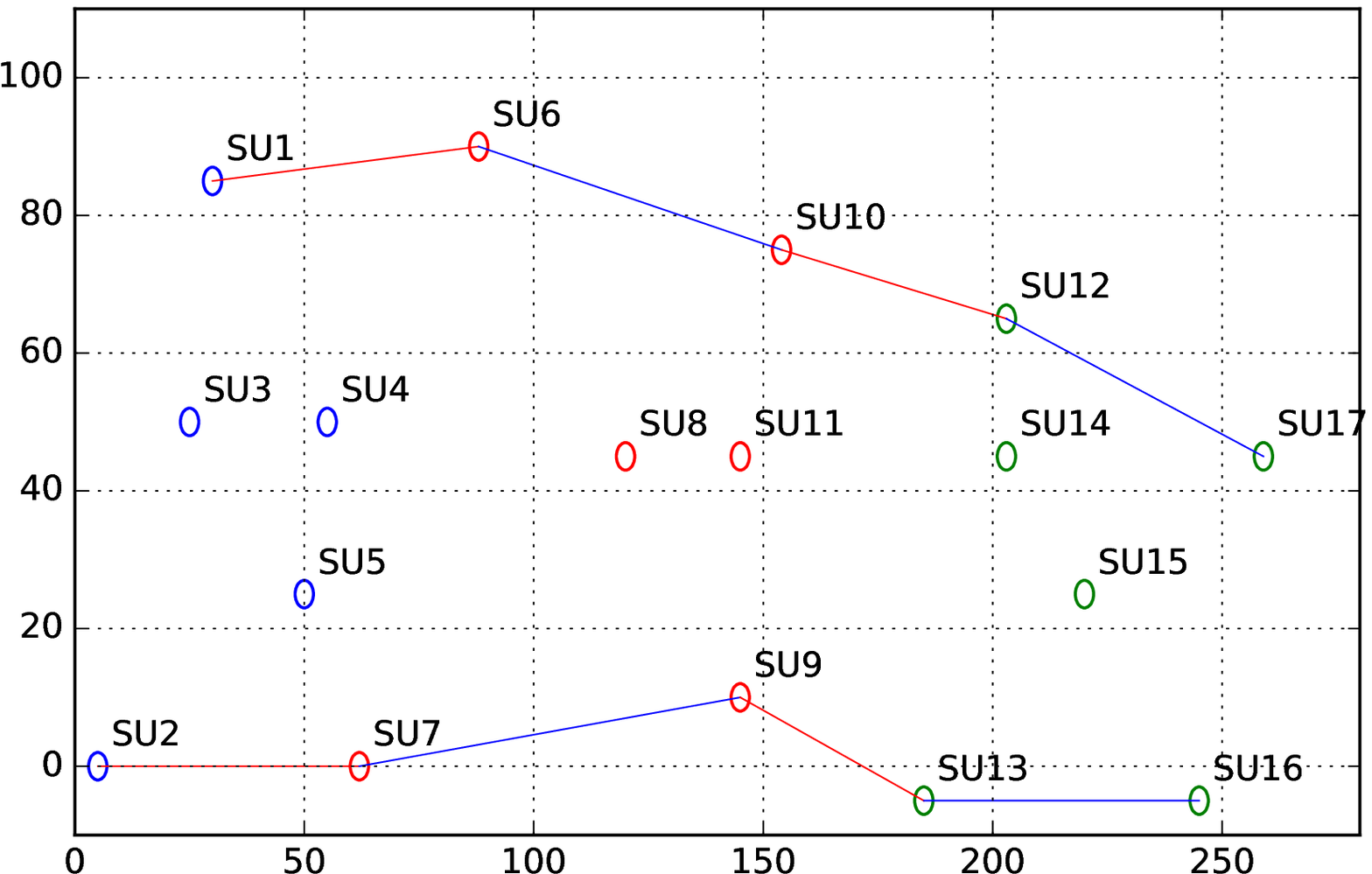}
  \caption{An attacker-free CRN over 2 PU channels. Red lines represent packet-forwarding over channel 1 with a higher probability. Blue lines represent 
  packet-forwarding over channel 0 with a higher probability.}
  \label{figure_best_route}
 \end{minipage}
 \begin{minipage}{.49\textwidth}
  \centering
  \includegraphics[width=1.0\textwidth]{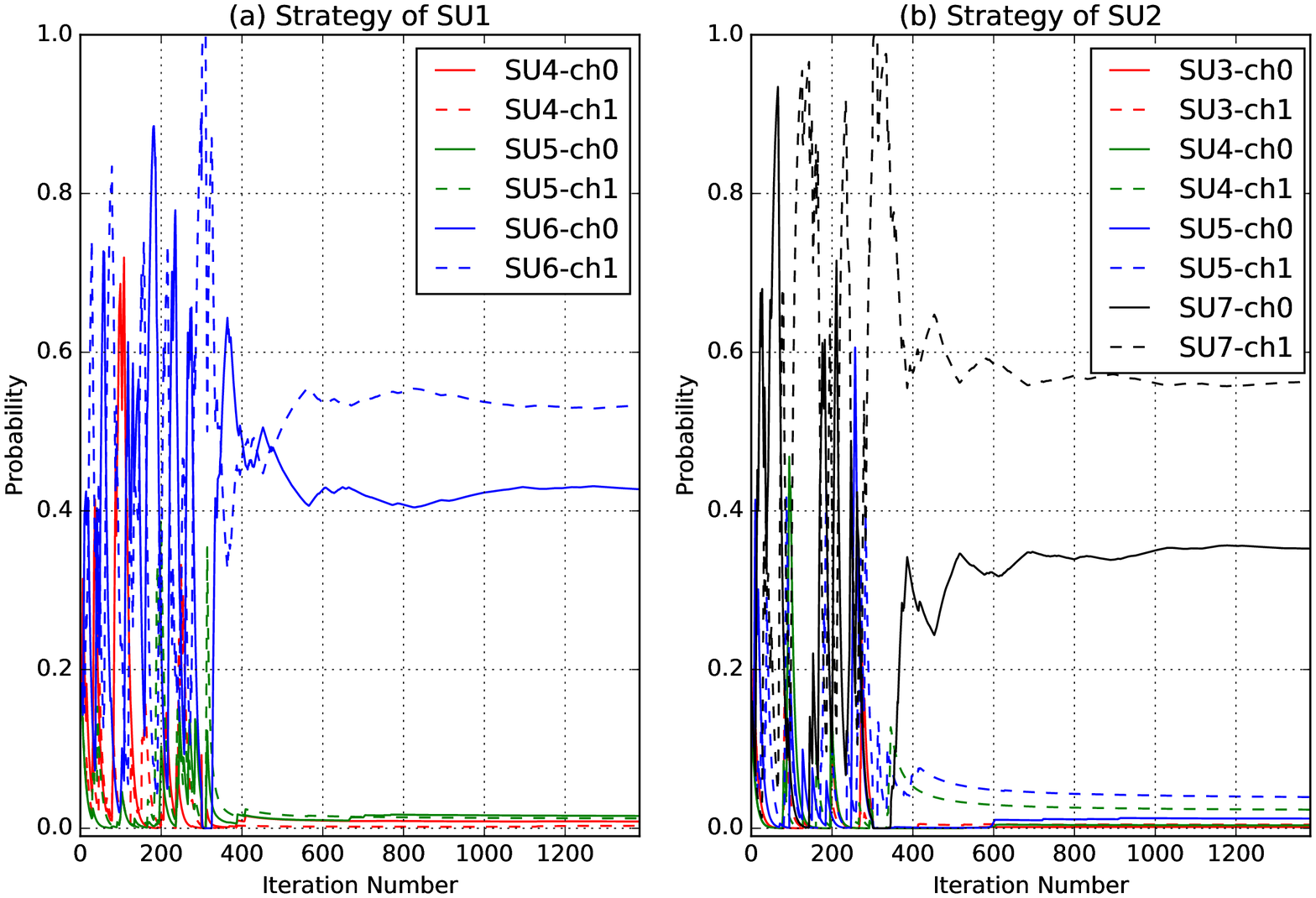}
  \caption{Strategy evolution: channel-relay selection probability vs. iteration number.}
  \label{figure_strategy_evolution}
 \end{minipage}
\end{figure}

Firstly, we demonstrate the convergence property of the proposed path selection mechanism given by (\ref{eq_action_value_local})-(\ref{eq_perturbed_br_modified}). 
Without loss of generality, we assume that the state transition maps are identical for all PU channels. We set the parameters for channel state transition
as $\lambda^{-1}\!=\!0.2$s, $\mu^{-1}\!=\!0.42$s, $T\!=\!0.5$s, and for a valid link $d^{\textrm{ETT}}=0.01$s.
For convenience of visualization, we examine a randomly generated 2-channel, 3-cluster CRN with 2 flows in Figure~\ref{figure_best_route}. 
In Figure \ref{figure_best_route}, SUs 1 and 2 are the source nodes and SUs 16 and 17 are the sink nodes. The strategy evolution for the source SUs is shown 
in Figure \ref{figure_strategy_evolution}. According to our discussion about channel contention on (\ref{eq_aloha_action_sender})-(\ref{eq_num_rts}),
any source selecting SUs 3, 4 or 5 as its relay will result in a higher probability of conflict with the other source. Therefore, SUs 1 
and 2 are expected to geographically separate their next hop as much as possible. As shown in Figure~\ref{figure_strategy_evolution}, with the learning scheme given by 
(\ref{eq_action_value_local})-(\ref{eq_perturbed_br_modified}), SUs 1 and 2 separate the two flows by choosing SUs 6 and 7 as their relays with non-zero 
probabilities. The strategies of relaying through SUs 3, 4 and 5 finally converge to near 0. A mixed-strategy NE
is reached and SUs 1 and 2 select between the two channels for transmission with non-zero probability. The highest-probability result of joint relay-channel selection   
 for each SU at the NE is shown by the colored lines in Figure~\ref{figure_best_route}.

\begin{figure}
  \centering
  \includegraphics[width=0.45\textwidth]{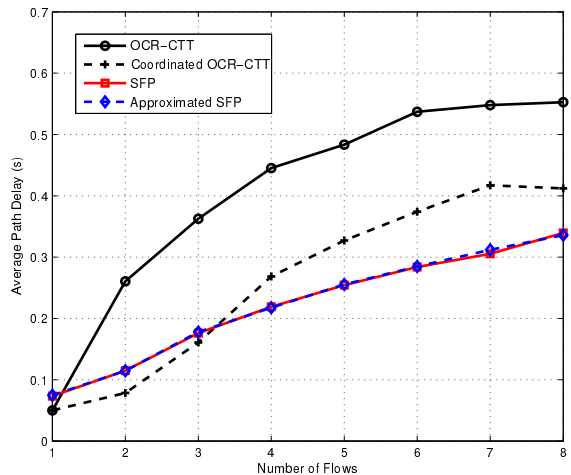}
  \caption{Average path delay vs. number of flows for different algorithms.}
  \label{figure_routing_comparison}
\end{figure}

In Figure~\ref{figure_routing_comparison}, we compare the performance of the algorithms given by (\ref{eq_action_value})-(\ref{eq_perturbed_br})
and (\ref{eq_action_value_local})-(\ref{eq_perturbed_br_modified}) with that of a reference algorithm based on Opportunistic Cognitive Routing (OCR) with 
Cognitive Transport Throughput (CTT) as the link performance metric \cite{6331686}. The original OCR-CTT algorithm was designed as a 
heuristic joint channel-relay searching method for efficient single-flow routing in CRNs. To address the bottleneck effect with multiple flows, we modify the 
original OCR-CTT algorithm by introducing a centralized, greedy channel assignment mechanism. The simulation is set in a $250$m$\times250$m area with 100 relays randomly 
deployed in a 2-channel, 3-cluster CRN. The coverage radius of each SU is set to 35m. As shown in Figure~\ref{figure_routing_comparison}, the proposed algorithms (SFP and Approximated SFP) with 
mixed-strategies have slightly larger delay than that of the deterministic OCR-CTT algorithms when the number of flows is small and the active SUs are sparse in the network. 
However, as the network becomes more congested with a larger number of flows, the proposed algorithms are able to better avoid channel conflicts and reduce the averge 
path delay by 30\% compared with the coordinated OCR-CTT algorithm.

\begin{figure}
 \begin{minipage}{.5\textwidth}
  \centering
  \includegraphics[width=0.90\textwidth]{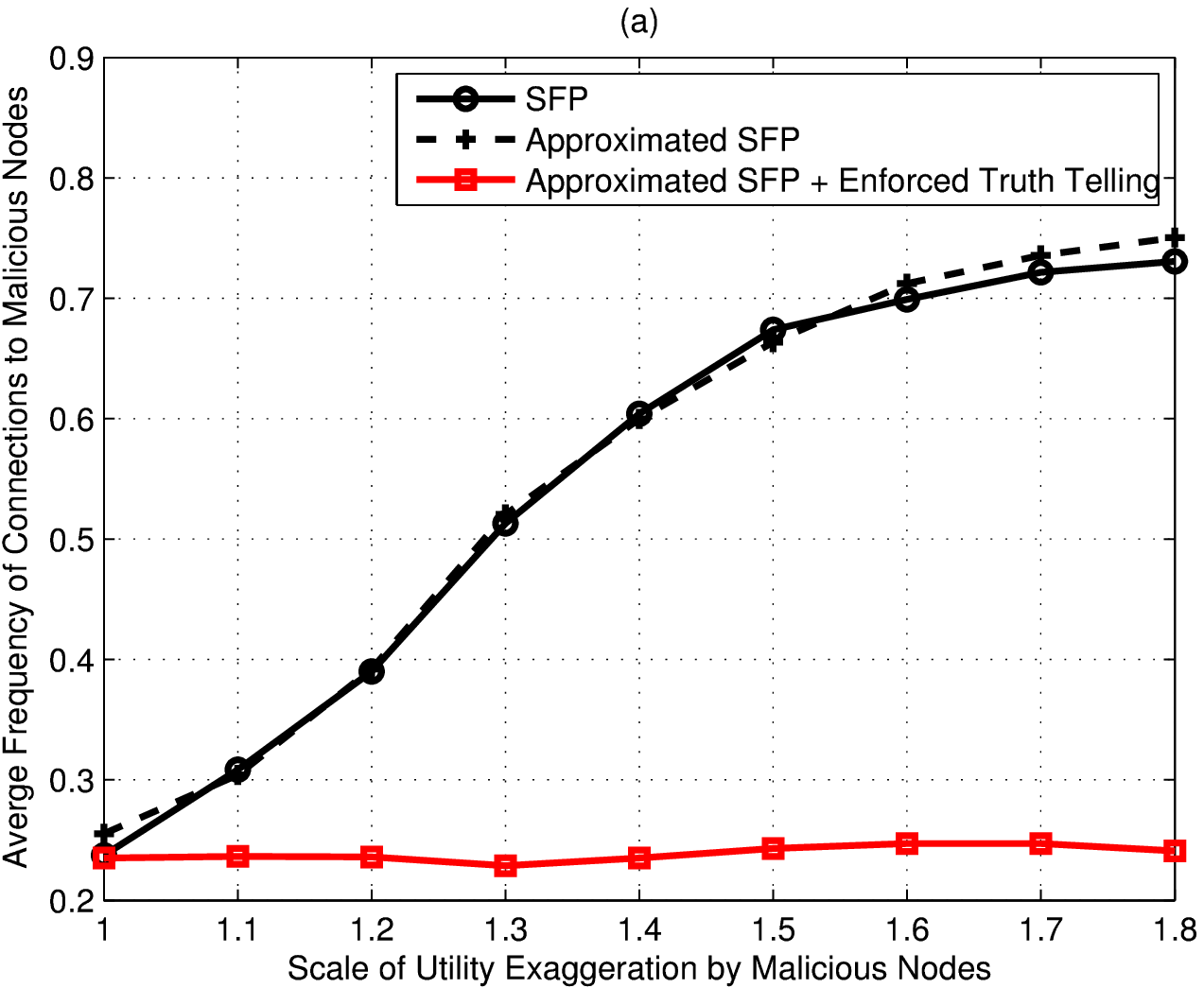}
 \end{minipage}
 \begin{minipage}{.5\textwidth}
  \centering
  \includegraphics[width=0.90\textwidth]{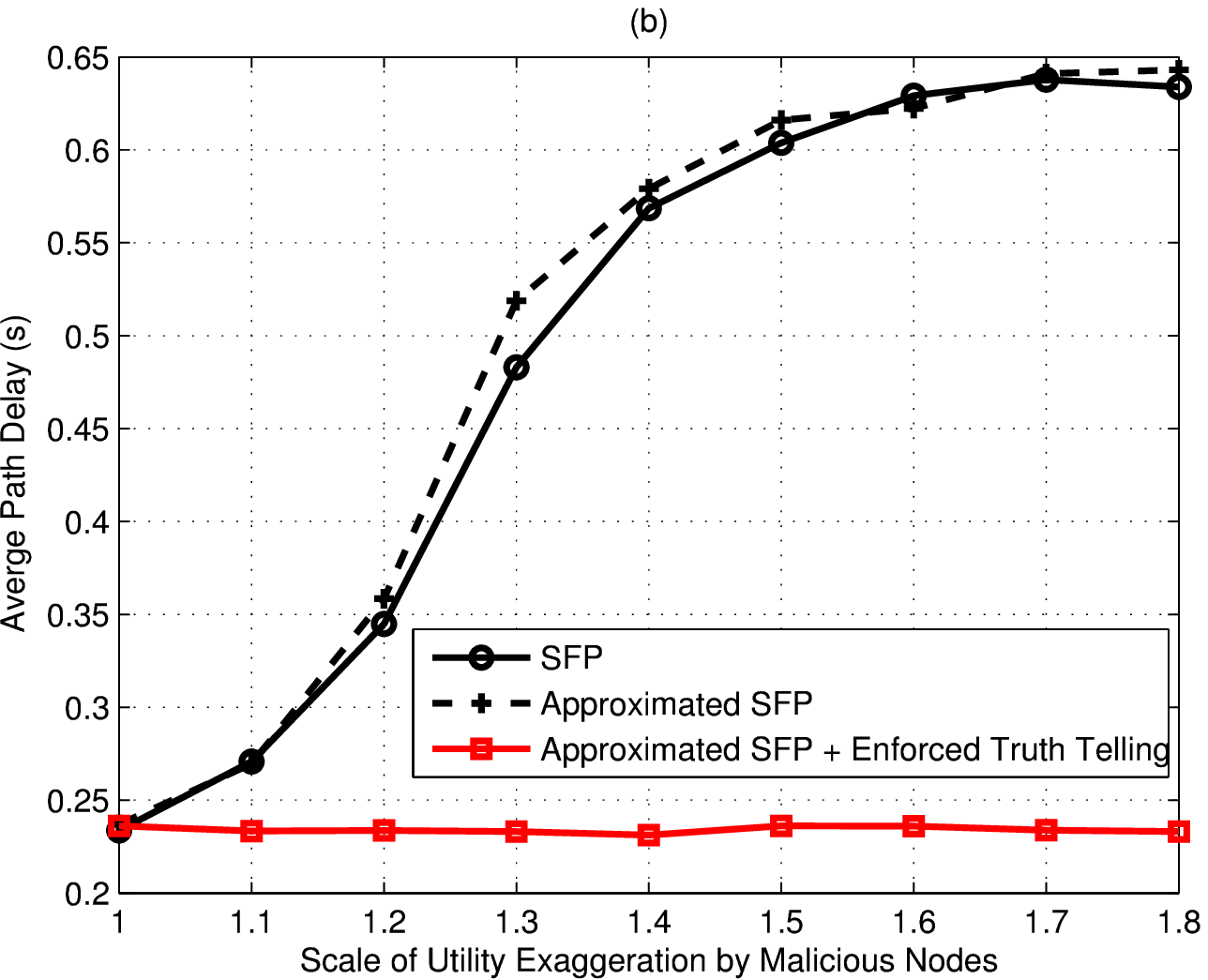}
 \end{minipage}
  \caption{(a) Frequency of connections to malicious nodes vs. scale for utility. (b) Average path delay vs. scale of exaggerated utility by malicious nodes.}
   \label{figure_attack_SH}
\end{figure}

In Figure~\ref{figure_attack_SH}, we evaluate the performance of the proposed strategy learning algorithm when malicious SUs exist. The simulation is conducted in the same 
randomly generated network for the simulation in Figure~\ref{figure_routing_comparison}. We investigate the ``aggressiveness'' of an attacker by varying 
the scale of information distortion by the malicious SUs based on the real value of the sub-path utility. The larger the scale that an attacker uses for information distortion, 
the more aggressive the attacker is. There are 4 flows in the CRN and for each source node there is one malicious SU randomly placed in its one-hop neighborhood.
Comparing the average path delay at 4 flows in Figure~\ref{figure_routing_comparison} with the average path delay at scale 1 in Figure~\ref{figure_attack_SH}, we note that 
the routing performance is not affected by the presence of attackers when malicious SUs do not adopt SH schemes. Intuitively, this is because with the proposed learning 
mechanisms, an SU is able to switch to alternative normal relays when performance deterioration from the attackers is detected and the network is not congested.
However, when truth-telling enforcement is not enabled, the malicious SUs are able to quickly attract the nearby flows by exaggerating their reported value of sub-path utility 
(see Figure \ref{figure_attack_SH}a). Consequently, a steep increase in average path delay can be observed in Figure \ref{figure_attack_SH}b. In contrast, when truth-telling enforcement 
is enabled, the performance of multi-flow routing remains in the same level of the case of no attackers. As can be observed in (\ref{eq_sigma}), given sufficient time for 
delay-evaluation based on the proposed feedback mechanism, the exponential operator in (\ref{eq_sigma}) is able to reduce the weight of non-optimal relays in 
(\ref{eq_perturbed_br_adjusted}) to near-zero. Therefore, as long as the network is not congested, the source node can only get connected to the malicious nodes if the routing
performance through the malicious nodes is no worse than the performance through any other neighbor nodes.

\vspace{-10.5pt}
\section{Conclusion}
\label{sec_conclusion}
In this paper, we have proposed a stochastic learning scheme for spectrum-aware, joint relay-channel selection in a multi-channel, multi-hop CRN. To address the potential
vulnerabilities due to the combined Routing-toward-Primary-User (RPU) and Sink-Hole (SH) attack, we have formulated the distributed routing process as a stochastic game.
By showing that the stochastic routing game can be decomposed into a group of single-state repeated games, we have proposed a a Stochastic Fictitious Play (SFP) based relay selection 
algorithm based on limited information back propagation. We have also introduced a Multi-Arm Bandit (MAB) based truth-telling enforcement procedure for normal SUs to evaluate 
the trustworthiness of their candidate relays. With numerical simulations, we have demonstrated that the proposed routing algorithm is able to reduce the average path delay 
by more than 30\% compared to conventional routing mechanisms. Moreover, we have demonstrated that with the proposed learning algorithm, it is guaranteed that the routing 
performance is not affected by the inside attackers.
\vspace{-10.5pt}
\appendix
\vspace{-10.5pt}
\subsection{Proof of Theorem \ref{thm_BEV}}
\label{app_thm_BEV}
Let $\mathcal{G}\!=\!\langle \mathcal{N}, \mathcal{S}, \mathcal{A}_i, \{r_i\}_{i\in{\mathcal{N}}}P(\mathbf{s}'|\mathbf{s},\mathbf{a})\rangle$ represents a general-case average-reward
recurrent stochastic game, where $r_i\!:\!{\mathcal{S}}\!\times\!{\mathcal{A}}\!\rightarrow\!\mathbb{R}$ ($\mathcal{A}\!=\!\times\mathcal{A}_i$) is bounded and state 
transition probability $P(\cdot)$ is a function of all the players' joint action $\mathbf{a}$. Let $R_i$ denote the expected average gain of player $i$ given in 
(\ref{eq_strategy_util}) and $g_i$ denote its expected bias value as in (\ref{eq_bv}). Then, for game $\mathcal{G}$ we have
\vspace{-8.5pt}
\begin{Lemma}[Theorem 2.6 of \cite{1997Altman}]
 \label{lemma_bv_aep}
 The joint strategy $\pmb\pi^*$ is an average NE point iff the pair of $R_i(\mathbf{s}, \pmb\pi^*)$ and $g_i(\mathbf{s}, \pmb\pi^*)$ solves the following optimality 
 equations for each play $i$:
  \begin{equation}
  \label{eq_lemma_bv_aep_1}
  \begin{array}{ll}
  R_i(\mathbf{s}, \pmb\pi^*)=\max\limits_{\pmb\pi_i}\Big\{\sum\limits_{\mathbf{s}'}P(\mathbf{s}'|\mathbf{s},\pmb\pi_i, \pmb\pi_{-i}^*)R_i(\mathbf{s}', \pmb\pi^*)\Big\},
  \end{array}
 \end{equation}
 \vspace{-4mm}
 \begin{equation}
  \label{eq_lemma_bv_aep_2}
  \begin{array}{ll}
  g_i(\mathbf{s}, \pmb\pi^*)=\max\limits_{\pmb\pi_i}\Big\{
  u_i(\mathbf{s}, \pmb\pi_i,\pmb\pi^*_{-i})-R_i(\mathbf{s}, \pmb\pi^*)+
  \displaystyle\sum\limits_{\mathbf{s}'}P(\mathbf{s}'|\mathbf{s},\pmb\pi_i, \pmb\pi_{-i}^*)g_i(\mathbf{s}', \pmb\pi^*)\Big\}.
  \end{array}
 \end{equation}
\end{Lemma}

According to Proposition \ref{prop_mdp}, the state transition in game $\mathcal{G}_r$ is independent of SU actions. Then, we readily obtain the two inequalities in
(i) of Theorem \ref{thm_BEV} according to (\ref{eq_lemma_bv_aep_2}).

To prove (ii) in Theorem \ref{thm_BEV}, we first consider the case of a normal SU. Based on Lemma \ref{lemma1}, we can substitute $h_{\mathcal{P}_i}(\mathbf{o},\pmb\pi^*)$ in 
(\ref{neq_bev_normal}) with (\ref{eq_lemma1}) and obtain $\forall \pmb\pi_i$:
\begin{equation}
 \label{eq_lemma_bv_aep_3}
 \begin{array}{ll}
 u_{\mathcal{P}_i}(\mathbf{o},\pmb\pi^*)\!-\!U_{\mathcal{P}_i}(\mathbf{o},\pmb\pi^*)\!+\!\displaystyle\sum\limits_{\mathbf{o}'}P(\mathbf{o}'|\mathbf{o})h_{\mathcal{P}_i}
 (\mathbf{o}',\pmb\pi^*)\!\ge\!
  u_{\mathcal{P}_i}(\mathbf{o},\pmb\pi_i,\pmb\pi_{-i}^*)\!-\!U_{\mathcal{P}_i}(\mathbf{o},\pmb\pi^*)\!+\!\displaystyle\sum\limits_{\mathbf{o}'}
   P(\mathbf{o}'|\mathbf{o})h_{\mathcal{P}_i}(\mathbf{o}',\pmb\pi^*).
 \end{array}
\end{equation}
From (\ref{eq_lemma_bv_aep_3}) we obtain 
  $u_{\mathcal{P}_i}(\mathbf{o},\pmb\pi^*)\!\ge\!u_{\mathcal{P}_i}(\mathbf{o},\pmb\pi_i,\pmb\pi_{-i}^*), \forall \pmb\pi_i$,
which is exactly the same as the condition equation for an NE in game $\mathcal{G}_r(\mathbf{o})$. For a malicious SU $j$, we can show $u_{\mathcal{P}_j}(\mathbf{o},
\pmb\pi^*)\le u_{\mathcal{P}_j}(\mathbf{o},\pmb\pi_j,\pmb\pi_{-j}^*)$ similarly with the help of (\ref{neq_bev_malicious}) in Theorem \ref{thm_BEV}. 

To show that the NE strategies for the stage game group $\mathcal{G}_r(\mathbf{o}:\forall \mathbf{o}\in\mathcal{O})$ constitute an NE strategy for $\mathcal{G}_r$, we rewrite 
(\ref{eq_lemma_prov_1}) as follows:
\begin{equation}
 \label{eq_lemma_bv_aep_5}
 U_{\mathcal{P}_i}(\mathbf{o},\pmb\pi)=\lim\limits_{\tau\rightarrow\infty}\frac{1}{\tau}\left(\sum_{n=0}^{\tau-1}P(\mathbf{o}'|\mathbf{o})u_{\mathcal{P}_i}(\mathbf{o}',\pmb\pi)\right).
\end{equation}
Consider the case that $\pmb\pi$ comprises of the NE strategies of the stage game groups, $\pmb\pi=(\pmb\pi^*(\mathbf{o}):\forall \mathbf{o}\in\mathcal{O})$. If $\pmb\pi$ is not an 
NE strategy of game $\mathcal{G}_r$, according to Definition \ref{Def_NE}, we can find at least one SU $i$ (assume that SU $i$ is normal), satisfying the following inequality:
\begin{equation}
 \label{neq_lemma_bv_aep_1}
 U_{\mathcal{P}_i}(\mathbf{o},\pmb\pi)-U_{\mathcal{P}_i}(\mathbf{o},\pmb{\tilde\pi}_i, \pmb\pi_{-i})<0, \exists\pmb{\tilde\pi}_i.
\end{equation}
Then, after substituting $U_{\mathcal{P}_i}$ in (\ref{neq_lemma_bv_aep_1}) with (\ref{eq_lemma_bv_aep_5}), we have:
\begin{equation}
 \label{neq_lemma_bv_aep_2}
 \begin{array}{ll}
 \displaystyle\lim\limits_{\tau\rightarrow\infty}\frac{1}{\tau}\Big(\sum_{n=0}^{\tau-1}P(\mathbf{o}'|\mathbf{o})\times
 \Big(u_{\mathcal{P}_i}(\mathbf{o}',\pmb\pi^*(\mathbf{o}'))- u_{\mathcal{P}_i}(\mathbf{o}',\pmb{\tilde\pi}_i(\mathbf{o}'), \pmb\pi^*_{-i}(\mathbf{o}'))\Big)\Big)<0,
 \end{array}
\end{equation}
which contradicts the fact that $\pmb\pi^*(\mathbf{o})$ is the NE strategy of stage game $\mathcal{G}_r(\mathbf{o})$. Therefore, property (ii) of Theorem 
\ref{thm_BEV} holds.
\vspace{-10.5pt}
\subsection{Proof of Theorem \ref{thm_layered_mdp}}
\label{app_thm_layered_mdp}
After exchanging the order of expectation and summation, we can expand (\ref{eq_strategy_util}) as:
\begin{equation}
  \label{eq_apdx_a_1}
  \begin{array}{ll}
  U_{\mathcal{P}_i}(\mathbf{o},\pmb\pi)=\displaystyle\lim\limits_{\tau\rightarrow\infty}\frac{1}{\tau}\sum_{n=0}^{\tau-1}E_{\mathbf{o}}\Big(u_i(\mathbf{o}(n),\pmb\pi)
  \Big\vert\mathbf{o}(0)\!=\!\mathbf{o}\Big)+\\
  E_{\pmb\pi_{i,1}}\!\Big\{\!\displaystyle\lim\limits_{\tau\rightarrow\infty}\!\frac{1}{\tau}\!\sum_{n=0}^{\tau-1}\!E_{\mathbf{o}}\Big(\!u_{\mathcal{P}_{a_{i,1}}}(\mathbf{o}(n),
  a_{i,1}, \pmb\pi_{i,2}, \pmb\pi_{-i})\Big\vert\mathbf{o}(0)\!=\!\mathbf{o}\!\Big)\!\Big\}
  \!=\!U_i(\mathbf{o},\pmb\pi)\!+\!E_{\pmb\pi_{i,1}}\Big\{U_{\mathcal{P}_{a_{i,1}}}(\mathbf{o},a_{i,1}, \pmb\pi_{i,2}, \pmb\pi_{-i})\Big\},
  \end{array}
\end{equation}
where $\pmb\pi_{i,1}$ is SU $i$'s strategy for choosing the next-hop SU. From (\ref{eq_bv}) and (\ref{eq_apdx_a_1}), we obtain:
\begin{equation}
 \label{eq_apdx_a_2} 
 \begin{array}{ll}
  h_{\mathcal{P}_i}(\mathbf{o}, \pmb\pi)=
  \displaystyle\lim\limits_{\tau\rightarrow\infty}\!\sum_{n=0}^{\tau-1}\!E_{\mathbf{o}}\!\bigg\{\!u_i(\mathbf{o}(n), \pmb\pi)\!+\!  
  E_{\pmb\pi_{i,1}}\!\left\{\!u_{\mathcal{P}_{a_{i,1}}}(\mathbf{o}(n), a_{i,1}, \pmb\pi_{i,2}, \pmb\pi_{-i})\!\right\}\\
  -\!U_i(\mathbf{o},\pmb\pi)\!-\!E_{\pmb\pi_{i,1}}\left\{U_{\mathcal{P}_{a_{i,1}}}(\mathbf{o},a_{i,1}, \pmb\pi_{i,2},\pmb\pi_{-i}))\right\}\Big|\mathbf{o}(0)=\mathbf{o}\bigg\}\\
  =\displaystyle\lim\limits_{\tau\rightarrow\infty}\sum_{n=0}^{\tau-1}E_{\mathbf{o}}\bigg\{u_i(\mathbf{o}(n), \pmb\pi)-U_i(\mathbf{o},\pmb\pi)\Big|\mathbf{o}(0)=\mathbf{o}\bigg\}
  +
  \displaystyle\lim\limits_{\tau\!\rightarrow\!\infty}\sum_{n=0}^{\tau-1}E_{\mathbf{o},\pmb\pi_{i,1}}\bigg\{
  u_{\mathcal{P}_{a_{i,1}}}(\mathbf{o}(n), a_{i,1}, \pmb\pi_{i,2}, \pmb\pi_{-i}))\\
  -U_{\mathcal{P}_{a_{i,1}}}(\mathbf{o},a_{i,1}, \pmb\pi_{i,2}, \pmb\pi_{-i}))\Big|\mathbf{o}(0)\!=\!\mathbf{o}\bigg\}
  =h_i(\mathbf{o}, \pmb\pi)+E_{\pmb\pi_{i,1}}\left\{h_{\mathcal{P}_{a_{i,1}}}(\mathbf{o}, a_{i,1}, \pmb\pi_{i,2}, \pmb\pi_{-i})\right\}.
  \end{array}
\end{equation}
Adding (\ref{eq_apdx_a_1}) and (\ref{eq_apdx_a_2}), we obtain:
\begin{equation}
  \label{eq_apdx_a_3} 
  \begin{array}{ll}
  h_{\mathcal{P}_i}(\mathbf{o},\pmb\pi)+U_{\mathcal{P}_i}(\mathbf{o},\pmb\pi)=h_{i}(\mathbf{o},\pmb\pi)+U_{i}(\mathbf{o},\pmb\pi)+\\
  E_{\pmb\pi_{i,1}}\!\left\{h_{\mathcal{P}_{a_{i,1}}}\!(\mathbf{o},a_{i,1},\pmb\pi_{i,2},\pmb\pi_{-i})\!+\!U_{\mathcal{P}_{a_{i,1}}}\!(\mathbf{o},a_{i,1},\pmb\pi_{i,2},\pmb\pi_{-i})
  \right\}.
  \end{array}
\end{equation}
After applying Lemma \ref{lemma1} to (\ref{eq_apdx_a_3}), (\ref{eq_layered_mdp}) is obtained.

Consider a normal SU $i\!\in\!\mathcal{N}$. From (\ref{eq_lemma1}), we can show that the best response of SU $i$ to the joint strategy $\tilde{\pmb\pi}_{-i}$ with respect to the
sum of its bias value and gain value is obtained when
\begin{equation}
 \label{eq_equivalence_solution}
 \begin{array}{ll}
  h_{\mathcal{P}_i}(\mathbf{o}, \tilde{\pmb\pi})+U_{\mathcal{P}_i}(\mathbf{o},  \tilde{\pmb\pi})
  =\max\limits_{\pmb\pi_i}\Big(u_{\mathcal{P}_i}(\mathbf{o}, \pmb\pi_i, \tilde{\pmb\pi}_{-i})
  +\sum\limits_{\mathbf{o}'}P(\mathbf{o}'|\mathbf{o})h_{\mathcal{P}_i}(\mathbf{o}, \tilde{\pmb\pi})\Big),
 \end{array}
\end{equation}
where $\tilde{\pmb\pi}$ is the solution to the right-hand side of (\ref{eq_equivalence_solution}). From (\ref{eq_lemma_bv_aep_1}) and (\ref{eq_lemma_bv_aep_2}) in Lemma 
\ref{lemma_bv_aep}, we have 
\begin{equation}
 \label{eq_equivalence_solution_2}
 \begin{array}{ll}
  h_{\mathcal{P}_i}(\mathbf{o}, \pmb\pi^*)+\max\limits_{\pmb\delta}U_{\mathcal{P}_i}(\mathbf{o},  \pmb\delta, \pmb\pi^*_{-i})
  =\max\limits_{\pmb\pi_i}\Big(u_{\mathcal{P}_i}(\mathbf{o}, \pmb\pi_i, \pmb\pi^*_{-i})
  +\sum\limits_{\mathbf{o}'}P(\mathbf{o}'|\mathbf{o})h_{\mathcal{P}_i}(\mathbf{o}, \pmb\pi^*)\Big),
 \end{array}
\end{equation}
and $\pmb\pi^*=(\pmb\delta^*,\pmb\pi^*_{-i})$ is the NE strategy. For the malicious SUs, a similar pair of equations to (\ref{eq_equivalence_solution}) and 
(\ref{eq_equivalence_solution_2}) can be obtained by substituting operator $\max(\cdot)$ with $\min(\cdot)$ in (\ref{eq_equivalence_solution}) and (\ref{eq_equivalence_solution_2}).
Comparing the right-hand side of (\ref{eq_equivalence_solution}) and (\ref{eq_equivalence_solution_2}), it is 
straightforward to show that the best response with respective to $h_{\mathcal{P}_i}(\mathbf{o}, {\pmb\pi})+U_{\mathcal{P}_i}(\mathbf{o},  {\pmb\pi})$ is also the 
NE strategy of the game.

\vspace{-10.5pt}
\subsection{Proof of Theorem \ref{thm_pseudotrajectory}}
\label{app_thm_Pseudotraj}
From \cite{leslie2003convergent}, we introduce Lemma \ref{lemma_pseduotrajectory} in regard to the two timescale learning process in (\ref{eq_action_value})-(\ref{eq_perturbed_br}):
\vspace{-8.5pt}
\begin{Lemma}[Theorem 5 of \cite{leslie2003convergent}]
 \label{lemma_pseduotrajectory}
  Consider that in the following stochastic approximation processes
  \begin{equation}
    \left\{
    \begin{array}{ll}
      \theta_1^{n+1}=\theta_1^{n}+\gamma^n_1(F_1(\theta_1^{n},\theta_2^{n})+M^{n+1}_1),\\
      \theta_2^{n+1}=\theta_2^{n}+\gamma^n_2(F_2(\theta_1^{n},\theta_2^{n})+M^{n+1}_2),
    \end{array}\right.
    \label{eq_lemma_two_processes}
  \end{equation}
for each $i$, $\theta_i^{n}$ is bounded, $\sum_{n\rightarrow\infty}\gamma^n_i=\infty$, $\sum_{n\rightarrow\infty}(\gamma^n_i)^2<\infty$, $F_i$ is globally Lipschitz continuous, 
$\{\sum^k_{n=1}\gamma^n_iM^{n}_i\}_k$ converges almost surely, and $\lim_{n\rightarrow\infty}\gamma^n_1/\gamma^n_2=0$. Suppose that for each $\theta_1$ the Ordinary Differential 
Equation (ODE)
\begin{equation}
 \label{eq_ODE_Y}
 \frac{\mathrm{d}Y}{\mathrm{d}t}=F_2(\theta_1, Y), \nonumber 
\end{equation}
has a unique globally asymptotically stable equilibrium point $\xi(\theta_1)$ such that $\xi$ is Lipschitz continuous. Then almost surely, 
\begin{equation}
 \label{eq_ODE_Y_pseudo_traj}
 \lim\limits_{n\rightarrow\infty}\Vert\theta_2^n-\xi(\theta_1^n)\Vert=0, \nonumber 
\end{equation}
and a suitable interpolation of the process $\{\theta_1^n\}$ is an asymptotic pseudo-trajectory of the flow defined by the ODE
\begin{equation}
 \label{eq_ODE_X}
 \frac{\mathrm{d}X}{\mathrm{d}t}=F_1(X, \xi(X)). \nonumber 
\end{equation}
\end{Lemma}

Let $\{\tilde{u}^n_{\mathcal{P}_i}(\mathbf{o},\mathbf{a}_i)\}$ in (\ref{eq_action_value}) be $\{\theta^n_2\}$ in (\ref{eq_lemma_two_processes}) and 
$\{\tilde{\pmb\pi}^n_i(\mathbf{o})\}$ in (\ref{eq_strategy_learning}) be $\{\theta^n_1\}$ in (\ref{eq_lemma_two_processes}), then we define the following two ODEs:
\begin{equation}
 \label{eq_def_ode_fast}
 \frac{\mathbf{d}\tilde{u}_{\mathcal{P}_i}(\mathbf{o}, \mathbf{a}_i)}{\mathbf{d}t}\!=\!F_2(\tilde{u}_{\mathcal{P}_i},\tilde{\pmb\pi}_i)\!=\!u_{\mathcal{P}_i}(\mathbf{o}, 
 \mathbf{a}_i)\!-\!\tilde{u}_{\mathcal{P}_i}(\mathbf{o},\mathbf{a}_{\mathcal{P}_i}),
\end{equation}
\begin{equation}
  \label{eq_def_ode_slow}
 \frac{\mathbf{d}\tilde{\pmb\pi}_i(\mathbf{o}, \mathbf{a}_i)}{\mathbf{d}t}\!=\!F_1(\mathbf{u}_{\mathcal{P}_i},\tilde{\pmb\pi}_i)\!=\!{\mathrm{BR}}(\mathbf{u}_{\mathcal{P}_i}
 (\mathbf{o}))\!-\!\tilde{\pmb\pi}_i(\mathbf{o},\mathbf{a}_i).
\end{equation}
According to our discussion on (\ref{eq_action_value}), $\tilde{u}_{\mathcal{P}_i}(\mathbf{o}, \mathbf{a}_i)$ almost surely converges to ${u}_{\mathcal{P}_i}(\mathbf{o}, 
\mathbf{a}_i, \pmb\pi_{-i})$. Then, by Lemma \ref{lemma_pseduotrajectory}, a suitable interpolation of $\{\tilde{\pmb\pi}^n_i(\mathbf{o})\}$ is an asymptotic 
pseudo-trajectory of the flow defined by the ODE in (\ref{eq_def_ode_slow}). It is well known \cite{leslie2003convergent, ECTA:ECTA376} that (\ref{eq_def_ode_slow}) is equivalent
to (\ref{eq_ODE_br}):
 \begin{equation}
  \label{eq_ODE_br}
  \frac{\mathrm{d}\pmb\pi_i(\mathbf{o})}{\mathrm{d}t}=\overline{\textrm{BR}}({\pmb\pi_{-i}}(\mathbf{o}))-\pmb\pi_i(\mathbf{o}).
 \end{equation}
where for a normal SU $i$ (we omit the state indicator $\mathbf{o}$ for simplicity)
\begin{equation}
 \label{eq_SFP_1}
 \begin{array}{ll}
 \overline{\textrm{BR}}({\pmb\pi_{-i}})\!=\!
 \arg\max\limits_{\pmb\pi_i}\!
 \Big(u_{\mathcal{P}_i}(\pmb\pi_i,\pmb\pi_{-i})\!-\!\lambda_i\sum\limits_{\mathbf{a}_i}\pmb\pi_i(\mathbf{a}_i)\log\pmb\pi_i(\mathbf{a}_i)\Big),
 \end{array}
\end{equation}
and for a malicious SU $j$
\begin{equation}
 \label{eq_SFP_2}
 \begin{array}{ll}
 \overline{\textrm{BR}}({\pmb\pi_{-j}})\!=\!
 \arg\!\max\limits_{\pmb\pi_j}\!
 \Big(\!u^{-1}_{\mathcal{P}_j}(\pmb\pi_j,\pmb\pi_{-\!j})\!-\!\lambda_j\!\displaystyle\sum\limits_{\mathbf{a}_j}\!\pmb\pi_j(\mathbf{a}_j)\!\log\pmb\pi_j(\mathbf{a}_j)\!\Big),\!
 \end{array}\!
\end{equation}
because (\ref{eq_perturbed_br}) provides the solutions to (\ref{eq_SFP_1}) and (\ref{eq_SFP_2}) \cite{ECTA:ECTA376}. In (\ref{eq_SFP_1}) and (\ref{eq_SFP_2}), the entropy function
$v_i(\pmb\pi_i)=-\sum_{\mathbf{a}_i}\pmb\pi_i(\mathbf{a}_i)\log\pmb\pi_i(\mathbf{a}_i)$ is called the perturbation in SFP. According to \cite{ECTA:ECTA376}, we have
\vspace{-8.5pt}
\begin{Lemma}[Proposition 3.1 of \cite{ECTA:ECTA376}]
 \label{lemma_SFP_vs_NE}
 Consider a general, normal-form repeated game $\mathcal{G}\!=\!\langle\mathcal{N},\times_{i\in\mathcal{N}}\mathcal{A}_i, \{u_i\}_{i\in\mathcal{N}}\rangle$. Let $\hat{\pi}^n_i$ 
 be the fixed point of the SFP dynamic given by (\ref{eq_ODE_br}) with respect to a perturbation vector $\mathbf{v}^n=(v_1^n,\ldots,v^n_{\vert\mathcal{N}\vert})$. If the perturbation 
 sequence $\{\mathbf{v}^n\}$ converges weakly, and the sequence $\{\hat{\pi}^n_i\}$ converges to ${\pi}^*_i$, then ${\pi}^*_i$ is the NE for $\mathcal{G}$.
\end{Lemma}
By Lemma \ref{lemma_SFP_vs_NE}, when the solution to the ODE in (\ref{eq_ODE_br}) converges to a fixed point, it converges to the NE of game $\mathcal{G}_r(\mathbf{o})$. Then,
based on the discussion following Lemma \ref{lemma_pseduotrajectory}, proving the convergence of the learning process given by (\ref{eq_action_value})-(\ref{eq_perturbed_br}) to 
the NE is equivalent to proving that the solution trajectories to the SFP dynamic in (\ref{eq_def_ode_slow}) converge to the set of fixed points from any initial condition. 
Observing the structure of 
$\mathcal{G}_r(\mathbf{o})$, the proof can be developed using the following properties of a repeated game:
\vspace{-8.5pt}
\begin{Lemma}[Corollary 5.5 of \cite{ECTA:ECTA376}]
 \label{le_convergence_supermodular}
 If a generic repeated game $\mathcal{G}$ is a supermodular game, then the solutions to the smooth best response dynamic in the form of (\ref{eq_ODE_br}) for $\mathcal{G}$ 
 converges almost surely to its rest point set from any initial condition. The remaining nonconvergent initial conditions are contained in a finite or countable union 
 $\cup_iM_i$, of invariant manifolds of codimension 1, and hence have measure zero.
\end{Lemma}
\vspace{-8.5pt}
\begin{Lemma}[Supermodular game\cite{han2012game}]
 \label{le_being_supermodular}
 A continuous normal-form game $\mathcal{G}={\langle}\mathcal{N}, \{\pmb{\Pi}_i\}_{i\in\mathcal{N}}, \{u_i(\pmb\pi_i)\}_{i\in\mathcal{N}}{\rangle}$ is a supermodular game if for 
 any player $i\in\mathcal{N}$,
 \begin{itemize}
  \item[i)] the strategy space $\pmb{\Pi}_i$ is a compact subset of $\mathbb{R}^K$.
  \item[ii)] the payoff function $u_i$ is upper semi-continuous in $\pmb\pi_i=(\pmb\pi_i,\pmb\pi_{-i})$.
  \item[iii)] $\frac{\partial^2u_i(\pmb\pi)}{\partial{\pmb\pi_{i,k}}\partial{\pmb\pi_{j,l}}}\ge0$  $\forall j\ne i, k, l$, where $\pmb\pi_{i,k}$ is the $k$-th element of vector
  $\pmb\pi_i$.
 \end{itemize}
\end{Lemma}

With Lemma \ref{le_being_supermodular}, we can check the supermodularity of game $\mathcal{G}_r(\mathbf{o})$ with respect to strategy $\pmb\pi_i$.  According to 
(\ref{eq_local_util}), we have $u_{i}\!\ge\!0$ $\forall
\mathbf{o}, \mathbf{a}$. Then, according to (\ref{eq_lemma_prov_2}), $\forall i\ne j$
\begin{equation}
 \label{two_order_diff}
 \frac{\partial^2(u_{\mathcal{P}_i}=\sum_{i\in\mathcal{P}_i}u_i(\pmb\pi))}{\partial{\pmb\pi_i(\mathbf{a}_i)}\partial{\pmb\pi_j}(\mathbf{a}_j)}\ge0, \forall \mathbf{a}_i,
 \mathbf{a}_j.
\end{equation}
Therefore, game $\mathcal{G}_r(\mathbf{o})$ in the form of continuous game\footnote{Such property also holds for malicious SUs as long as their strategy
learning scheme complies with SFP given by (\ref{eq_perturbed_br}).} with strategy 
$\pmb\pi$ is a supermodular game. By Lemma \ref{le_convergence_supermodular}, the smooth best response dynamic converges almost surely. By Lemma \ref{lemma_SFP_vs_NE}, Theorem 
\ref{thm_pseudotrajectory} is proved.

\vspace{-10.5pt}
\subsection{Proof of Theorem \ref{thm_convergence_approximation}}
\label{app_thm_convergence_approximation}
The proof of Theorem \ref{thm_convergence_approximation} can be achieved by applying Lemma \ref{lemma_pseduotrajectory} repeatedly to the learning scheme given by 
(\ref{eq_action_value_local}) and (\ref{eq_action_value_path}), then to the learning scheme given by (\ref{eq_action_value_path}) and (\ref{eq_strategy_learning_modified}). 
According to our discussion on (\ref{eq_action_value_local}), $\tilde{u}_i^n(\mathbf{o},\mathbf{a}_i)$ has a unique globally asymptotically stable equilibrium ${u}_i(\mathbf{o},
\mathbf{a}_i, \pmb\pi_{-i})$ if $\pmb\pi$ is fixed. Then, it is sufficient to prove that the following ODE:
\begin{eqnarray}
\label{app_thm_conv_1}
\begin{array}{ll}
 \displaystyle\frac{\mathbf{d}\tilde{u}_{\mathcal{P}_i}\!(\mathbf{o})}{\mathbf{d}t}\!=\!\!
 \Big(\!\displaystyle\sum_{\mathbf{a}_i}\!\tilde{\pmb\pi}_i(\mathbf{o},\mathbf{a}_i)\!
  (\tilde{u}_{i}(\mathbf{o},\mathbf{a}_i)\!+\!\tilde{u}_{\mathcal{P}_{\mathbf{a}_{i,1}}}(\mathbf{o}))\!-\!\tilde{u}_{\mathcal{P}_i}\!(\mathbf{o})\!\Big),
  \end{array}
\end{eqnarray}
is globally asymptotically stable to show that the learning process given by (\ref{eq_action_value_local})-(\ref{eq_strategy_learning_modified}) produces an asymptotic 
pseudo-trajectory of the SFP flow. Omitting state indicator $\mathbf{o}$ for convenience, we denote $\hat{u}_{\mathcal{P}_i}(\mathbf{a}_i)\!=\!\tilde{u}_{i}(\mathbf{a}_i)\!+
\!\tilde{u}_{\mathcal{P}_{\mathbf{a}_{i,1}}}$, $\xi_i\!=\!\frac{\mathbf{d}\tilde{u}_{\mathcal{P}_i}}{\mathbf{d}t}$ and 
$\epsilon(\mathbf{a}_i)\!=\!\frac{\mathbf{d}\hat{u}_{\mathcal{P}_i}(\mathbf{a}_i)}{\mathbf{d}t}$, and define a Lyapunov function:
\begin{eqnarray}
\label{app_thm_conv_2}
\begin{array}{ll}
 V_i(t)=
 \Big(\displaystyle\sum_{\mathbf{a}_i}\tilde{\pmb\pi}_i(\mathbf{a}_i)
  \big(\tilde{u}_{i}(\mathbf{a}_i)\!+\!\tilde{u}_{\mathcal{P}_{\mathbf{a}_{i,1}}}\big)\!-\!\tilde{u}_{\mathcal{P}_i}\Big)^2
  \end{array}.
\end{eqnarray}
We sort the SUs in path $\mathcal{P}_i$ according to their distance in hop count to sink $L$ in an ascending order as $\{L-1, L-2, \ldots, i\}$.
Then, the two-timescale stochastic
approximation process in Lemma \ref{lemma_pseduotrajectory} can be extended to multiple-timescale with the same form of function $F_i$ as in (\ref{eq_lemma_two_processes}):
\begin{equation}
\label{eq_multi_scale}
\left\{
\begin{array}{ll}
 F_{1}(\tilde{u}_j(\mathbf{a}_j), \tilde{\pmb\pi})={u}_j(\mathbf{a}_j(n))-\tilde{u}^n_j(\mathbf{a}_j),\\
 F^j_{2}(\tilde{u}_j(\mathbf{a}_j), \tilde{u}_{\mathcal{P}_{j}}, \tilde{u}_{\mathcal{P}_{j\!+\!1}})\!=\!
 \displaystyle\sum_{\mathbf{a}_i}\!\tilde{\pmb\pi}_i(\mathbf{a}_i)
  (\tilde{u}_{i}(\mathbf{a}_i)\!+\!\tilde{u}_{\mathcal{P}_{\mathbf{a}_{j}}})\!-\!\tilde{u}_{\mathcal{P}_i}.
  \end{array}\right.
\end{equation}

Since the learning process in (\ref{eq_action_value_local}) is globally asymptotically convergent, then at the stable point of $\tilde{u}_i$, 
$\frac{\mathbf{d}\tilde{u}_i}{\mathbf{d}t}\!=\!0$ and $\epsilon_i(\mathbf{a}_i)\!=\!\frac{\mathbf{d}\tilde{u}_{\mathcal{P}_{i+1}}}{\mathbf{d}t}$, where $\mathbf{a}_{i,1}\!=\!i\!+\!1$.
We now examine $V_i$ and obtain
\begin{equation}
\label{app_thm_conv_3}
\begin{array}{ll}
 \displaystyle\frac{1}{2}\frac{\mathbf{d}V_i}{\mathbf{d}t}=
 \Big(\displaystyle\sum_{\mathbf{a}_i}\tilde{\pmb\pi}_i(\mathbf{a}_i)
  \hat{u}_{\mathcal{P}_i}(\mathbf{a}_i)\!-\!\tilde{u}_{\mathcal{P}_i}\Big)\times
 \left(\displaystyle\frac{\mathbf{d}}{\mathbf{d}t}\sum_{\mathbf{a}_i}\frac{e^{\lambda_i\hat{u}_{\mathcal{P}_i}(\mathbf{a}_i)}}{\sum_{\mathbf{b}}e^{\lambda_i\hat{u}_{\mathcal{P}_i}
  (\mathbf{b})}}\hat{u}_{\mathcal{P}_i}(\mathbf{a}_i)-\frac{\mathbf{d}\tilde{u}_{\mathcal{P}_i}}{\mathbf{d}t}\right),\\
  =\!\xi_i\!\bigg(\!\displaystyle\sum_{\mathbf{a}_i}\bigg(\!\sum_{\mathbf{b}}\frac{\lambda_ie^{\lambda_i\hat{u}_{\mathcal{P}_i}(\mathbf{a}_i)
  \hat{u}_{\mathcal{P}_i}(\mathbf{b})}}{\left(\sum_{\mathbf{b}}e^{\lambda_i\hat{u}_{\mathcal{P}_i}(\mathbf{b})}\right)^2}(\epsilon_i(\mathbf{a}_i)\!-\!\epsilon_i(\mathbf{b}))
  \hat{u}_{\mathcal{P}_i}(\mathbf{a}_i)
  +\displaystyle\frac{e^{\lambda_i\hat{u}_{\mathcal{P}_i}(\mathbf{a}_i)}}{\sum_{\mathbf{b}}e^{\lambda_i\hat{u}_{\mathcal{P}_i}(\mathbf{b})}}\epsilon_i(\mathbf{a}_i)\bigg)
  -\xi_i\bigg),\\
  =\!\lambda_i\!\displaystyle\sum_{\mathbf{a}_i}\!\sum_{\mathbf{b}}\frac{e^{\lambda_i\hat{u}_{\mathcal{P}_i}(\mathbf{a}_i)}}{\sum_{\mathbf{b}}e^{\lambda_i
  \hat{u}_{\mathcal{P}_i}(\mathbf{b})}}(\epsilon_i(\mathbf{a}_i)\!-\!\epsilon_i(\mathbf{b}))\hat{u}_{\mathcal{P}_i}\!(\mathbf{a}_i)\xi_i
  +\displaystyle\sum_{\mathbf{a}_i}\frac{e^{\lambda_i\hat{u}_{\mathcal{P}_i}(\mathbf{a}_i)}}{\sum_{\mathbf{b}}e^{\lambda_i\hat{u}_{\mathcal{P}_i}(\mathbf{b})}}
  \epsilon_i(\mathbf{a}_i)\xi_i\!-\!\xi^2_i.
  \end{array}\nonumber
\end{equation}

We start examining the property of $\frac{\mathbf{d}V_i}{\mathbf{d}t}$ in the way of backward propagation from SU $L-1$. Since $\tilde{u}_{\mathcal{P}_{L}}\!=\!0$, we have
$\epsilon_{L\!-\!1}(\mathbf{a}_{L\!-\!1})\!=\!0$, hence $\frac{1}{2}\frac{\mathbf{d}V_{L\!-\!1}}{\mathbf{d}t}\!=\!-\xi^2_{L\!-\!1}\!\le\!0$ at the stable point of the approximation process 
represented by $F^{L-1}_2(\tilde{u}_{L-1},\tilde{\pmb\pi})$. Therefore, the ODE for SU $L-1$ in the form of (\ref{app_thm_conv_1}) is globally asymptotically convergent. Then, we can 
apply Lemma \ref{lemma_pseduotrajectory} to the two-timescale learning process featured by $F^{L-1}_2$ and $F^{L-2}_2$, and show that a suitable interpolation of the process 
$\{\tilde{u}^n_{\mathcal{P}_{L-2}}\}$ is an asymptotic pseudo-trajectory of the flow defined by the ODE $\frac{\mathbf{d}\tilde{u}_{\mathcal{P}_{L-2}}}{\mathbf{d}t}$ given in 
(\ref{app_thm_conv_1}). At the stable point of $\tilde{u}_{\mathcal{P}_{L-1}}$, we have $\frac{\mathbf{d}\tilde{u}_{\mathcal{P}_{L-1}}}{\mathbf{d}t}\!=\!0$, so 
$\epsilon_{L\!-\!2}(\mathbf{a}_{L-2})\!=\!0$. With the similar way to analyzing $\frac{\mathbf{d}V_{L-1}}{\mathbf{d}t}$, we have 
$\frac{1}{2}\frac{\mathbf{d}V_{L-2}}{\mathbf{d}t}\!=\!-
\xi^2_{L-2}\!\le\!0$. By repeatedly applying the same analysis to the sequence of the learning processes featured by $\{(F^{L-1}_2,F^{L-2}_2), (F^{L-2}_2,F^{L-3}_2),\ldots, 
(F^{i+1}_2,F^{i}_2)\}$, we obtain Lemma \ref{le_convergence_approximation}:
\vspace{-8.5pt}
\begin{Lemma}
 \label{le_convergence_approximation}
 The learning process given in (\ref{eq_action_value_local}) and (\ref{eq_action_value_path}) is globally asymptotically convergent, provided that the following are satisfied:
 $\lim\limits_{n\rightarrow\infty}\sum\limits_n\alpha(n)\!=\!\infty$, $\lim\limits_{n\rightarrow\infty}\sum\limits_n\alpha^2(n)\!<\!\infty$, 
 $\lim\limits_{n\rightarrow\infty}\sum\limits_n\gamma_i(n)\!=\!\infty$, $\lim\limits_{n\rightarrow\infty}\sum\limits_n\gamma_i^2(n)\!<\!\infty$, 
 $\lim\limits_{n\rightarrow\infty}(\gamma_i(n)/\alpha(n))\!=\!0$ and $\lim\limits_{n\rightarrow\infty}(\gamma_i(n)/\gamma_j(n))\!=\!0$, if SU $i$ is closer to the sink SU than 
 SU $j$ in terms of hop count. 
\end{Lemma}
If $\lim\limits_{n\rightarrow\infty}(\beta(n)/\gamma_i(n))\!=\!0$, we can further conclude that the learning process given by (\ref{eq_strategy_learning_modified}) and 
(\ref{eq_perturbed_br_modified}) yields an asymptotic pseudo-trajectory of the flow defined by the SPF-based ODE. We note that Lemma \ref{le_convergence_supermodular} and 
Lemma \ref{le_being_supermodular} still hold for a new game with the utility of each player being the convergent biased value estimation. Then, Theorem 
\ref{thm_convergence_approximation} is proved.
\vspace{-8.5pt}

%
\bibliographystyle{IEEEtran}
\bibliography{Reference}
\end{document}